\documentclass[usenatbib,usegraphicx,letterpaper]{mn2e}
\usepackage[totalwidth=480pt,totalheight=680pt]{geometry}

\usepackage{amssymb}
\usepackage{epsfig}
\usepackage{amsmath}

\newcommand{\beq}{\begin{equation}}
\newcommand{\eeq}{\end{equation}}
\newcommand{\ben}{\begin{enumerate}}
\newcommand{\een}{\end{enumerate}}
\newcommand{\bit}{\begin{itemize}}
\newcommand{\eit}{\end{itemize}}
\newcommand{\beqray}{\begin{eqnarray}}
\newcommand{\eeqray}{\end{eqnarray}}

\newcommand{\monetwo}{m_{12}}

\newcommand{\phirand}{\Phi_{\mathrm{rand}}}
\newcommand{\phiglo}{\Phi_{\mathrm{global}}}
\newcommand{\ffos}{F_{\mathrm{fos}}}
\newcommand{\lcen}{L_{\mathrm{cen}}}

\newcommand{\li}{L_{\mathrm{i}}}
\newcommand{\nj}{N_{\mathrm{j}}}
\newcommand{\phisat}{\Phi_{\mathrm{sat}}}
\newcommand{\phicen}{\Phi_{\mathrm{cen}}}

\newcommand{\vpeak}{V_{\mathrm{max}}^{\mathrm{peak}}}
\newcommand{\vmax}{V_{\mathrm{max}}}
\newcommand{\vacc}{V_{\mathrm{max}}^{\mathrm{acc}}}
\newcommand{\vzero}{V_{\mathrm{max}}^{\mathrm{z=0}}}
\newcommand{\vsub}{V_{\mathrm{sub}}}
\newcommand{\vl}{V_{\mathrm{L}}}
\newcommand{\ngal}{n_{g}}
\newcommand{\msun}{M_\odot}
\newcommand{\nh}{n_{h}}

\newcommand{\mri}{M_{\mathrm{r}}^{i}}
\newcommand{\mr}{M_{\mathrm{r}}}
\newcommand{\mh}{M_{\mathrm{h}}}
\newcommand{\sigmai}{\sigma^{i}}
\newcommand{\veff}{V_{\mathrm{eff}}}
\newcommand{\vbolshoi}{V_{\mathrm{Bolshoi}}}

\newcommand{\linit}{M^{\mathrm{init}}_{\mathrm{r}}}
\newcommand{\lscatter}{M^{\mathrm{scatter}}_{\mathrm{r}}}

\newcommand{\nrank}{n_{\mathrm{rank}}}

\newcommand{\Omegam}{\Omega_{M}}

\newcommand{\littleh}{h}

\newcommand{\tilt}{n_{s}}
\newcommand{\sigmaeight}{\sigma_{8}}
\newcommand{\lcdm}{\Lambda\mathrm{CDM}}

\newcommand{\hmpc}{h^{-1}\mathrm{Mpc}}

\newcommand{\ith}{i^{\mathrm{th}}}

\newcommand{\dd}{\mathrm{d}}

\begin{document}

\title[SHAM Beyond Clustering]
{SHAM Beyond Clustering: New Tests of Galaxy-Halo Abundance Matching with Galaxy Groups}

\author[A.P. Hearin et al.]
{Andrew P. Hearin$^{1,2}$,
Andrew R. Zentner$^2$,
Andreas A. Berlind$^3$, 
Jeffrey A. Newman$^2$ \\
$^1$ Fermilab Center for Particle Astrophysics, 
Fermi National Accelerator Laboratory, Batavia, Illinois 60510-0500; aphearin@fnal.gov \\
$^2$ Department of Physics and Astronomy \& Pittsburgh Particle Physics, Astrophysics and Cosmology Center (PITT PACC),\\ 
University of Pittsburgh, Pittsburgh, PA 15260; zentner@pitt.edu, janewman@pitt.edu\\
$^3$ Department of Physics and Astronomy, Vanderbilt University, Nashville, TN;
a.berlind@vanderbilt.edu
}

\maketitle

\begin{abstract} 
We construct mock catalogs of galaxy groups using subhalo abundance matching (SHAM) and undertake several new tests of the SHAM
prescription for the galaxy-dark matter connection. {\em All} SHAM models we studied exhibit significant tension with galaxy groups observed in the Sloan Digital Sky
Survey (SDSS). The SHAM prediction for the field galaxy luminosity function is systematically too dim, and the group galaxy luminosity function systematically too bright, regardless of the details of the SHAM prescription. 
SHAM models connecting r-band luminosity, $\mr,$ to $\vacc,$ the maximum circular velocity of a subhalo at the time of accretion onto the host, faithfully reproduce
the abundance of galaxy groups as a function of richness, $g(N).$ However, SHAM models connecting $\mr$ with $\vpeak,$ the peak value of $\vmax$ over the entire merger history of the halo, over-predict the abundance of galaxy groups.
Our results suggest that SHAM models for the galaxy-dark matter connection may be unable to simultaneously reproduce the observed group multiplicity function and two-point projected galaxy clustering. 
Nevertheless, we also report a new success of the abundance matching prescription: an accurate prediction for $\Phi(\monetwo),$ the abundance of galaxy groups as a
function of magnitude gap $\monetwo,$ defined as the difference
between the r-band absolute magnitude of the two brightest group
members. We demonstrate that it may be possible to use joint measurements of $g(N)$ and $\Phi(\monetwo)$ 
 to provide tight constraints on the details of the SHAM implementation.
Additionally, we show that the
hypothesis that the luminosity gap is constructed via random draws
from a universal luminosity function provides a poor description of
the data, contradicting recent claims in the literature. 
Finally, we test a common assumption of the Conditional Luminosity Function formalism, 
that the satellite luminosity function $\Phi_{\mathrm{sat}}(L)$ need only be conditioned by the brightness of the central galaxy $L_{\mathrm{cen}}.$ 
We find this assumption to be well-supported by the observed magnitude gap distribution.
\end{abstract}


\maketitle

%---------------------------
\section{Introduction}
\label{section:introduction} 
%--------------------------- 

The centers of dark matter halos are the natural sites for galaxy
formation, as these are the locations of the deepest gravitational
potential wells in the universe \cite[e.g.,][]{white_rees78}. 
The development of a theory of galaxy
formation that encompasses the complex array of physical processes
known to contribute to cosmic structure formation is one of the
fundamental goals of astrophysics, and ennumerating the connection between 
galaxies and dark matter halos may help to establish the foundations of any 
such theory. Additionally, such a connection can serve as 
an empirical link between large-scale survey data and theoretical 
predictions.  Our contemporary theory of cosmology, $\lcdm,$
makes precise, quantitative predictions for the distribution of dark
matter in the universe over a wide range of scales, and so establishing 
the galaxy-dark matter connection is a key step toward unlocking the
predictive power of $\lcdm.$

One of the most commonly used techniques for connecting dark matter
halos to galaxies is subhalo abundance matching (SHAM). The
fundamental tenet of all SHAM models is that there is a monotonic
mapping between some elementary property of galaxies (usually
luminosity or stellar mass) and an elementary property of halos. SHAM
models determine this mapping through the implicit relation defined by
matching the predicted abundance of halos with the observed abundance
of galaxies. When used in concert with numerical simulations of
cosmological structure formation, abundance matching techniques have
been shown to predict accurately galaxy clustering statistics
\citep{kravtsov_etal04,tasitsiomi_etal04,conroy_etal06,behroozi_etal10,moster_etal10},
the Tully-Fisher relation \citep{trujillo-gomez_etal11}, and the
conditional stellar mass function \citep{reddick_etal12}.  
To date, the vast majority of tests of the SHAM algorithm have 
been based upon clustering.  We will present new tests of  
SHAM based on statistics that are distinct from the widely quoted clustering tests.

In $\lcdm,$ gravitationally
self-bound structures form hierarchically, with tiny peaks in the
initial cosmic density field collapsing into very small dark matter
halos that gradually merge together to form groups and clusters of
galaxies. Galaxy groups are thus interesting environments for testing
theories of structure formation in general, and the galaxy-halo
connection in particular. Indeed, the influence of the group
environment on galaxy properties has a long history and has 
received considerable attention in the recent literature,
\cite[e.g.,][]{yang_etal05,zandivarez_etal06,robotham_etal06,yang_etal08,yang_etal09,blanton_berlind07,tinker_etal11a,gerke_etal12}. 

In this paper we investigate SHAM predictions for the assembly of
galaxies into groups and test these predictions against galaxy groups
observed in SDSS. 
%We conduct our study of abundance matching by first constructing a SHAM-based mock catalog of galaxies and then applying a group-finding algorithm on the mock catalog that is identical to that used to find groups in the SDSS galaxy sample. Details of the SHAM algorithm, such as the particular halo property used in the abundance matching, vary in the literature, and we have investigated several choices for the specific implementation of the SHAM algorithm with the aim of determining which methods best reproduce the observed properties of galaxy groups. 
By comparing properties of observed galaxy groups and those of a SHAM-based mock sample, we
provide a series of new tests of the abundance matching prescription
connecting galaxies to dark matter halos. 

One of the fundamental
properties of any catalog of groups is the {\em multiplicity
function}, the abundance of groups as a function of {\em richness},
$N,$ the number of members in a group. Previous studies of galaxy
group catalogs \citep{berlind_weinberg02,berlind_etal06,vale_ostriker06} have
demonstrated that measurements of the group multiplicity function
$g(N)$ may contain valuable information about how galaxies populate
dark matter halos. Motivated by this, we use the group multiplicity function as one of our primary, new tests of SHAM. 
As an additional group-based statistical test, we compare the SHAM prediction for the galaxy luminosity function
conditioned on whether or not the galaxies are members of a group.

%While the abundance matching algorithm permits an exact reproduction of the observed galaxy luminosity function by construction, SHAM need not necessarily reproduce the luminosity function of a subsample of galaxies that has been conditioned on some property. 

One galaxy group property that has received attention from a rapidly
growing body of literature is the luminosity gap, $\monetwo,$ the
difference in r-band absolute magnitude between the two most luminous
members of a galaxy group. Significant investigation has been focused on a
class of systems known as {\em fossil groups}, usually defined
as an X-ray bright ($L_{X,bol}>10^{42}$erg/s) group of galaxies with
$\monetwo \geq 2.$ The prevailing theoretical paradigm for 
fossil group formation is that these systems 
have evolved quiescently for a significant fraction
of a Hubble time, during which dynamical friction has had sufficient time to cause
the biggest satellites to merge with the central galaxy, resulting in
a massive, bright central galaxy with few bright satellites
\citep{ponman94,jones_etal03,donghia_etal05,donghia_etal07,dariush_etal07,vonbendabeckmann_etal08,milosavljevic_etal06,vikhlinin_etal99,miller_etal11,labarbera_etal12,tavasoli_etal11,aguerri_etal11}. 
For a recent paper on fossil groups that includes an excellent review of
the history of their study, we refer the interested reader to
\citet{harrison_etal12}. To test this paradigm within the SHAM framework, 
 we study
$\Phi(\monetwo),$ the abundance of our mock and observed groups as a
function of $\monetwo,$ finding that this statistic has the potential 
to constrain the manner in which galaxies populate dark
matter halos. This conclusion is consistent with previous work
\citep{skibba_etal07}, showing that the relative brightness of the
central galaxy in a group and its brightest satellite is influenced by
parameters governing the conditional luminosity function.

In a paper studying a very similar galaxy group catalog to the one we use here,
\citet{paranjape_sheth11} found that the observed $\Phi(\monetwo)$ is
consistent with the hypothesis that the brightness distribution of
galaxy group members is determined by a set of random draws from a
universal luminosity function. This finding implies that, for a galaxy
population of a given luminosity function, the gap abundance is
uniquely determined by knowledge of the abundance of groups as a
function of richness. This result is particularly surprising in light
of recent results \citep{hearin_etal12} demonstrating that the
magnitude gap contains information about group mass that is
independent of richness \citep[see also][]{ramella_etal07}. A possible resolution to this apparent
discrepancy was recently pointed out in \citet{more12}: the global gap
abundance $\Phi(\monetwo)$ is a mass function-weighted sum over the
mass-conditioned gap abundance, $\Phi(\monetwo|M),$ and so it is
possible in principle that the magnitude gap depends on both mass and richness in
such a way that the mass function-weighting washes out any
statistically-significant mass dependence in the global
$\Phi(\monetwo).$ 

In \S~\ref{section:mcs} we show that the global gap abundance
exhibited by galaxy groups in SDSS is inconsistent with the random
draw hypothesis, contradicting the claims in
\citet{paranjape_sheth11} that derive from their measurements of one-point statistics. 
As discussed in Appendix B, we find that the treatment of fiber collisions in the group catalog used by \citet{paranjape_sheth11}, together with the definition of the magnitude gap adopted in the \citet{paranjape_sheth11} measurement
of $\Phi(\monetwo),$ are responsible for the difference between our
conclusions. These findings are in keeping with the marked correlation function analysis appearing in \citet{paranjape_sheth11}, which also suggested that group galaxy brightnesses are inconsistent with the random draw hypothesis.
Additionally, we generalize these data randomization
techniques to conduct a direct test of a common assumption of the Conditional Luminosity Function (CLF) formalism, 
namely that the satellite luminosity function need only be
conditioned on the brightness of the central galaxy.  

%---------------------------
\subsection{Outline}
\label{subsection:outline}
%---------------------------

This paper is organized as follows. 
We briefly describe the SDSS catalog of galaxy groups against which we compare our predictions in \S~\ref{section:data}. 
We describe our methods for constructing our mock catalogs in \S~\ref{section:mocks}, and provide a detailed description of some novel features of our SHAM implementation in Appendix A. 
In \S~\ref{subsection:gn} we
compare the observed multiplicity function to that which is exhibited
by our mocks. In \S~\ref{subsection:groupfield} we present our predictions for and measurements of the group-membership-conditioned luminosity function. 
We test the abundance matching prediction for luminosity gap statistics in  \S~\ref{subsection:pgap}. 
Our tests of several random draw hypotheses appear in \S~\ref{section:mcs}; conclusions based on such statistics are sensitive to the treatment of fiber collisions, which we describe in detail in Appendix B.  
We discuss our results and
compare them to those in the existing literature in
\S~\ref{section:discussion}, and conclude with a brief overview of our
primary findings in \S~\ref{section:conclusion}. In Appendix C we demonstrate the robustness of our group multiplicity function results to possible systematic errors caused by edge effects and fiber collisions. 

%---------------------------
\section{Data}
\label{section:data}
%---------------------------

We study galaxy group properties in a volume-limited catalog of 
groups identified in Sloan Digital Sky Survey (SDSS) 
Data Release 7 \cite[][DR7 hereafter]{sdss_dr7}
using the algorithm described in \cite{berlind_etal06}. 
This catalog is an update of the \citet{berlind_etal06} group 
catalog (which was based on SDSS Data Release 3). 
The galaxies in this sample are all members of the Main
Galaxy Sample of SDSS DR7. Groups in this galaxy catalog are
identified via a redshift-space friends-of-friends algorithm that
takes no account of member galaxy properties beyond their redshifts and
positions on the sky. Our groups are constructed from galaxies in
a volume-limited spectroscopic sample ($\veff \simeq 5.8 \times10^{6}(\hmpc)^3$) in the redshift range 
$0.02 \le z \le 0.068$ with r-band absolute magnitude $M_r - 5\log h < -19$. 
We refer to this catalog as the ``Mr19'' group catalog. Each of the
$6439$ groups in the Mr19 catalog contains $N \geq 3$ members. We refer
the reader to \citet{berlind_etal06} for further details on the  group
finding algorithm. 

Fiber collisions occur when the angular separation between two or more
galaxies is closer than the minimum separation permitted by the plugging mechanism
 of the optical fibers used to measure galaxy spectra in SDSS 
\citep[see][and references therein]{guo_etal12}. We briefly note here that
fiber collisions in the DR7-based galaxy sample we use in this paper are treated differently than in the catalog presented in \citet{berlind_etal06}, which was based on DR3 data. As we will see in \S~\ref{section:mcs}, this different
treatment has important consequences for the measurement of magnitude
gaps. In Appendix B we discuss these differences in detail and argue
that the convention adopted in the DR3-based catalog induces systematic errors in magnitude gap
measurements that can be avoided if fiber collisions are instead
modeled as we do in this work. Throughout this paper, when we refer to the richness $N$ and the velocity dispersion $\sigma_v$ of a group, we include the fiber-collided members. When we compute the magnitude gap of a group (see \S~\ref{subsection:pgap}) and the luminosity function of galaxies in our sample (see \S~\ref{subsection:groupfield}), we exclude fiber-collided galaxies.

%---------------------------
\section{Mock Catalogs}
\label{section:mocks}
%---------------------------

We compare the SDSS DR7 group data to a mock catalog of 
galaxy groups based on the Bolshoi N-body simulation \citep{klypin_etal11}. 
The Bolshoi simulation models the cosmological growth of structure 
in a cubic volume $250\,\littleh^{-1}\mathrm{Mpc}$ on a side within a 
standard $\lcdm$ cosmology with total matter density $\Omegam=0.27$, 
Hubble constant $\littleh=0.7$, 
power spectrum tilt $\tilt=0.95$, 
and power spectrum normalization $\sigmaeight=0.82$. 
The Bolshoi data are available at {\tt http://www.multidark.org} and we refer the 
reader to \citet{riebe_etal11} for additional information.
Our analysis requires reliable identification of self-bound 
subhalos within the virial radii of distinct halos. We utilize the 
{\tt ROCKSTAR} \citep{behroozi_etal11} halo finder in order to identify 
halos and subhalos within Bolshoi. 

We utilize the subhalo abundance matching (SHAM) technique to associate 
galaxies with dark matter halos. Although abundance matching is widely 
used to construct mock galaxy catalogs 
\cite[e.g.,][]{kravtsov_etal04,tasitsiomi_etal04,conroy_etal06,watson_etal11,hearin_etal12,reddick_etal12}, 
our particular implementation of SHAM is novel and so we describe it 
in detail in Appendix A. In this section we provide a brief sketch of our 
SHAM prescription and review the primary advantages of our implementation.

SHAM models assume a monotonic relationship between the stellar masses 
of galaxies and the maximum circular speeds of test particles 
within their host dark matter halos, $\vmax\equiv\mathrm{max}\left[\sqrt{GM(<r)/r}\right]$, 
where $r$ is the distance from the halo center and $M(<r)$ is the halo mass 
contained within a distance $r$ of the halo center. 
 Inferring galaxy stellar masses is non-trivial, 
so in practice galaxy luminosities are often used to associate galaxies 
with halos using SHAM, though this may introduce important biases that are less prevalent for samples selected by stellar mass. For example, simulation studies suggest that 
satellite galaxies may be assigned halo masses that are biased low due to the fact that satellite galaxies  have older stellar populations than central galaxies  \cite[e.g.,][]{simha_etal12}.  
It is most common to construct SHAM assignments for SDSS data using the r-band absolute luminosities of the galaxies as rough proxies for stellar mass.  
We follow this approach in this paper and numerous previous authors have shown that SHAM based on r-band 
luminosity is surprisingly reliable to describe halo clustering \citep{kravtsov_etal04,conroy_etal06} and group statistics \citep{reddick_etal12} despite 
the fact that there are several effects that can lead to differing luminosities at fixed stellar mass. 

Assuming a monotonic relationship 
between $\vmax$ and r-band absolute magnitude $\mr,$ the SHAM 
galaxy-halo assignment follows by requiring the cumulative number density of halos with circular velocity $\vmax$ 
to be equal to the cumulative number density of the galaxies with brightness $\mr.$  
As a further complication, subhalos within 
host dark matter halos evolve significantly due to interactions within 
the dense environments of the larger host halos.  As a result, the present 
values of $\vmax$, which we denote $\vzero$, may be a poor proxy for stellar 
masses or r-band luminosities.  It is now common practice to assign luminosities 
to subhalos based on their values of $\vmax$ {\em evaluated at the time 
at which they merged into their distinct host halos}, $\vacc$ \citep{conroy_etal06}.  Often, 
a halo can be significantly affected by interactions prior to entering 
the virialized region of a distinct host halo, so it is also interesting 
to explore using the maximum value of $\vmax$ ever attained by a subhalo 
as the stellar mass/luminosity proxy, $\vpeak$ \citep{behroozi_etal12,reddick_etal12}.  

Tthe SHAM assignment of r-band luminosities to halos and subhalos occurs through 
the implicit relation 
\beq
\label{eq:lv}
\ngal(<M_r)=\nh(>\vl), 
\eeq
where $\ngal(<\mr)$ is the number density of observed galaxies with r-band 
magnitudes brighter than $\mr$ \citep{blanton_etal05}, 
and $\nh(>\vl)$ is the predicted number density of dark matter halos 
and subhalos with assigned circular speeds $>\vl$. 
As we mentioned in the previous paragraph, the circular speeds assigned 
to subhalos are often not their circular speeds at the time of 
observation.  Following this common practice, the quantity $\vl$ is conventionally evaluated as 
\begin{eqnarray}
\label{eq:vassign}
\quad \quad \vl & = & \vzero \quad \mathrm{(host\ halos)} \nonumber \\
    & = & \vsub   \quad \ \mathrm{(subhalos)} \nonumber
\end{eqnarray}
where $\vsub=\vzero$ if one chooses to use $\vzero$ to describe 
the luminosities of subhalos, $\vsub=\vacc$ if one chooses to 
use the maximum circular velocity at accretion for subhalos, 
and $\vsub=\vpeak$ if one chooses to use the maximum value of 
$\vmax$ ever attained by the subhalo.  
Throughout this paper, 
we refer to mock catalogs constructed in this fashion using $\vzero$ 
as ``SHAM0'' catalogs, those built with $\vacc$ as ``SHAMacc''. However, we follow a different convention for SHAM models constructed using $\vpeak$ as the abundance matching property. For $\vpeak$ models, which we label as ``SHAMpeak", {\em we use $\vpeak$ for both host halos and subhalos}, rather than only for the subhalos. Our reason for adopting this convention is because this is the SHAM method which \citet{reddick_etal12} found to best describe galaxy clustering and the conditional stellar mass function, and we wish to test this model specifically with the set of observables we study here.
We explore the predictions of each of these three models for several previously unexplored statistics describing the observed distribution of galaxies at low-redshift.

The advantages of SHAM-like models is that they are 
simple, they embody the fundamental theoretical prejudice that dark 
matter halos that represent deeper gravitational wells should host 
larger (more luminous) galaxies, and such models describe galaxy 
clustering over a range of redshifts remarkably well 
\cite[e.g.,][]{kravtsov_etal04,tasitsiomi_etal04,conroy_etal06,behroozi_etal12}.  Specifically, there is significant 
empirical support for SHAM models premised on 
$\vacc$ as well as $\vpeak.$ 
\citet{trujillo-gomez_etal11} showed that SHAMacc models provide an accurate prediction for the observed Tully-Fisher relation, and 
\cite[e.g.,][]{conroy_etal06} found that the projected two-point galaxy correlation function predicted by SHAMacc models is in good agreement with the SDSS measurements of projected galaxy clustering. However, a recent, more detailed analysis \citep{reddick_etal12} showed that SHAMpeak models provide more accurate predictions for clustering on small scales ($0.1\hmpc\lesssim r_{\mathrm{p}}\lesssim0.5\hmpc$).

In practice, some scatter between $\vmax$ and $\mr$ is often introduced into the 
basic SHAM assignments. The scatter accounts for the fact that galaxy formation 
is a complex process, so a single halo parameter cannot specify a stellar 
mass (or luminosity). Perhaps more importantly, scatter brings SHAM predictions 
into better agreement with galaxy clustering statistics 
\citep[e.g.,][]{tasitsiomi_etal04,behroozi_etal10,reddick_etal12}. 
Additionally, the observed Tully-Fisher relation has intrinsic scatter \citep{pizagno_etal07}, and so SHAM models without scatter cannot accurately describe the properties of the observed galaxy distribution.

We investigate 
the influence of scatter on our results using three different models of 
scatter between circular velocity and absolute magnitude. 
Our fiducial model, whose construction is described in detail in Appendix A, 
is designed to be similar to the model explored in \citet{trujillo-gomez_etal11}. At fixed $\vmax,$ our fiducial model
 has $0.2$dex of scatter in luminosity at the faint end 
($-19.5\lesssim\mr\lesssim-19$)
and $0.15$dex of scatter at the bright end
($-22\lesssim\mr\lesssim-21.5$).
However, we note that our model cannot be precisely the same as that of \citet{trujillo-gomez_etal11} 
because we use a different SHAM algorithm to incorporate scatter. 
Second, we explore a model which has a constant scatter of $0.1$dex.  
We will refer to this as our ``alternate'' scatter model. 
Third, we consider models with no scatter between $\mr$ and $\vmax.$

The SHAM construction summarized in Eq.~(\ref{eq:lv}) enables luminosities 
to be assigned to dark matter halos 
in a manner that can match {\em any} observed galaxy luminosity function. 
A detail of SHAM implementation is that a specific choice must be made 
for the galaxy luminosity function 
to use for the SHAM assignments.  A common and convenient 
choice is to use a fit to observed luminosity functions, 
such as that provided by \citet{blanton_etal05}.  
However, in order to ensure that our results are not affected by 
the residuals of any such fit, the mock catalogs that we 
construct match the global luminosity function of the 
Mr19 galaxies {\em exactly}.  Enforcing the requirement that the SHAM 
galaxy catalog have a luminosity function that matches the 
observed luminosity function exactly complicates the 
introduction of scatter into the SHAM prescription. 
To enforce the observed galaxy luminosity function 
on our SHAM assignments with scatter in the 
$\vl$-$M_r$ relation, we use a novel implementation 
of SHAM, the details of which are given in Appendix A.  
The important features of this implementation are as 
follows.
\ben
\item[1.] Our mock galaxy catalog has a luminosity function 
that, by construction, matches the observed Mr19 luminosity function {\em exactly}, 
even after including scatter.
\item[2.] The amount of scatter in the $\vl$-$M_r$ relation can be specified simply, so that 
implementing SHAM assignments with differing amounts of scatter is straightforward.
\item[3.] Even when the above two requirements are met, the algorithm is very fast, lending itself 
to applications that require the construction of a large number of mock catalogs.  This advantage 
is significant compared to, for example, the algorithm of \citet{trujillo-gomez_etal11}.
\een

Once galaxies with r-band luminosities have been assigned to dark matter halos and subhalos, 
we proceed to find groups as follows. First, the galaxies inherit the spatial positions of their host 
(sub)halos in the simulation. We use the distant observer approximation and make the z-axis of the simulation cube into the line-of-sight, and then use the (sub)halo mean velocities in the z-direction to move the mock galaxies into redshift space. At this point, each mock galaxy has a redshift coordinate and two spatial coordinates, and so we identify groups of mock galaxies using the same algorithm that was applied to the observational data. 
This guarantees that our mock groups are subject to the same redshift-space projection effects as 
the Mr19 catalog. 

Once groups have been found, we introduce fiber collisions to our mock galaxies as follows. 
For each mock group of richness $N$ we randomly select a Mr19 group with a similar richness. 
If the number of fiber-collided members of the randomly selected group is $N_{\mathrm{fc}},$ 
then we randomly choose $N_{\mathrm{fc}}$ of the members of the mock group and assign fiber 
collisions to these members. This procedure ensures that the fraction of fiber-collided galaxies 
in our mock groups {\em scales with richness in the same way as it does in the data}, a feature that is important for correct, detailed predictions of 
 the magnitude gap (see \S~\ref{subsection:pgap} and Appendix B). For all the statistics described in this paper, we include or exclude fiber-collided mock galaxies according to the same conventions that we do with the data, as described at the end of \S~\ref{section:data}.

Of course, the treatment described above does not account for the spatial biases of fiber collisions, but we 
do not focus on spatial clustering in the current study. Additionally, this modeling does not encode the influence of fiber collisions on group identification, which {\em is} relevant to our study of the group multiplicity function (see \S~\ref{subsection:gn}). However, as discussed in Appendix C, we find that the change to $g(N)$ induced by fiber collisions is sufficiently small that it does not influence any of our conclusions, and so our treatment of fiber collisions suits the aims of this paper.

We conclude this section by noting that none of the SHAM tests studied in this work include the possible influence of so-called {\em orphan galaxies}, defined as galaxies occupying a dark matter halo whose mass has dropped below the resolution limit of the simulation. We return to this point in \S~\ref{subsection:orphans}.

%
%---------------------------
\section{New Tests of Abundance Matching}
\label{section:prediction}
%---------------------------

In this section, we scrutinize the abundance matching prescription 
for the mapping between galaxies and dark matter halos in several novel ways.  
In \S~\ref{subsection:gn}, we consider the number density of groups 
as a function group richness, in \S~\ref{subsection:groupfield} we 
study the group and field luminosity functions of galaxies, and in 
\S~\ref{subsection:pgap} we investigate the distribution of galaxy 
luminosities within groups, concentrating significant attention on 
the magnitude gap statistic. As discussed in \S~\ref{section:mocks}, 
we explore three distinct assignments for the effective maximum 
circular velocities $\vmax$ to be used in the luminosity assignments of 
subhalos. Briefly, we refer to mocks based on $\vmax$ at the time accretion as ``SHAMacc'', mocks based on the redshift-zero value of $\vmax$ as ``SHAM0,'' and mocks based on the peak value of $\vmax$ over the entire merger history of the halo as ``SHAMpeak.''  
We also analyze SHAM predictions based upon three distinct models for 
the amount of scatter between $M_r$ and circular velocity.  
Throughout this paper we refer to the SHAMacc model 
with $0.2$dex of scatter at the faint end and $0.15$dex of 
scatter at the bright end, following \citet{trujillo-gomez_etal11}, 
as our default model.  We refer explicitly to results from the other 
models where relevant.

%---------------------------
\subsection{Multiplicity Function}
\label{subsection:gn}
%---------------------------

%---------------------------------------------------------------------------------------------------
\begin{figure*}
\centering
\includegraphics[width=8cm]{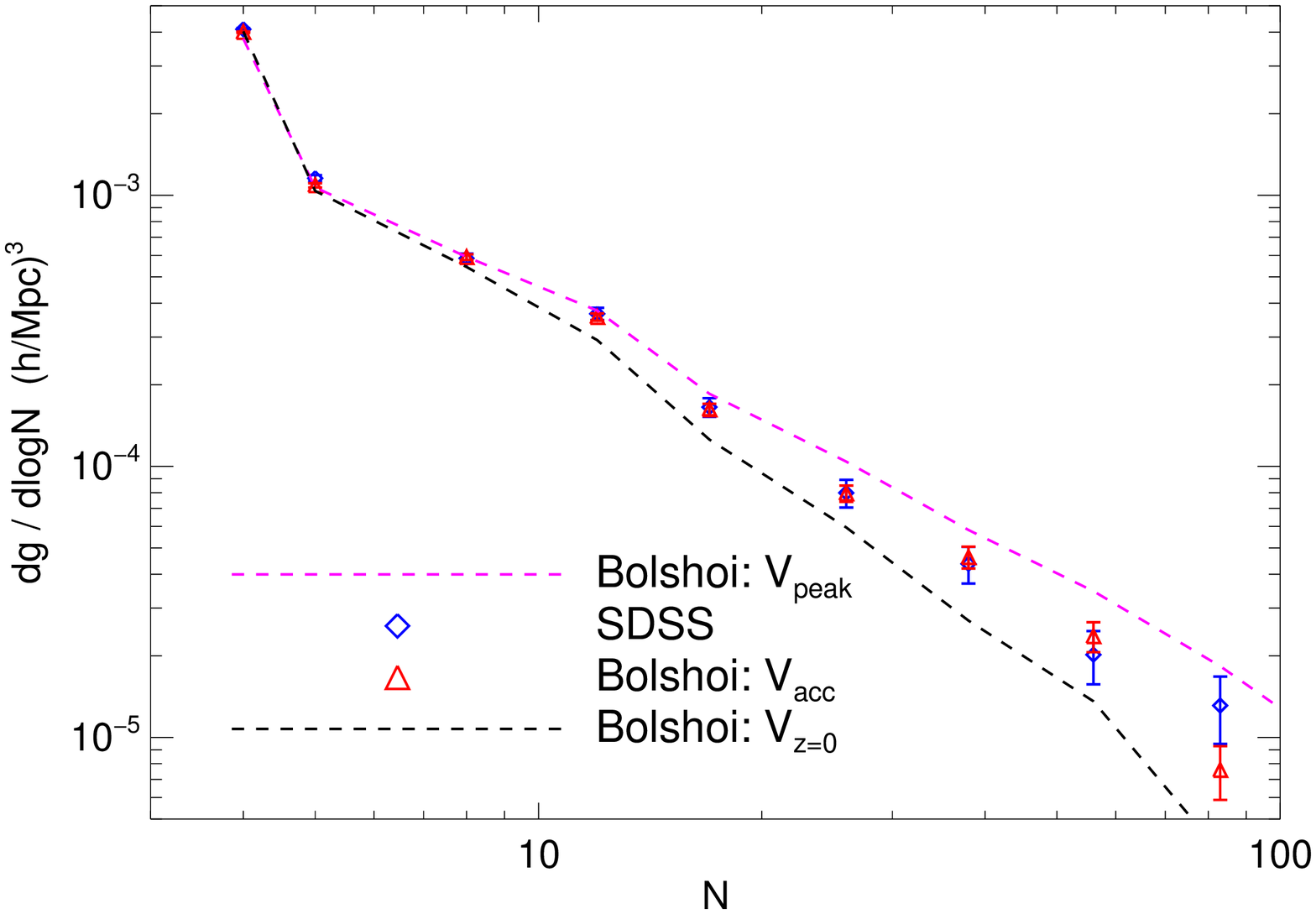}
\includegraphics[width=8cm]{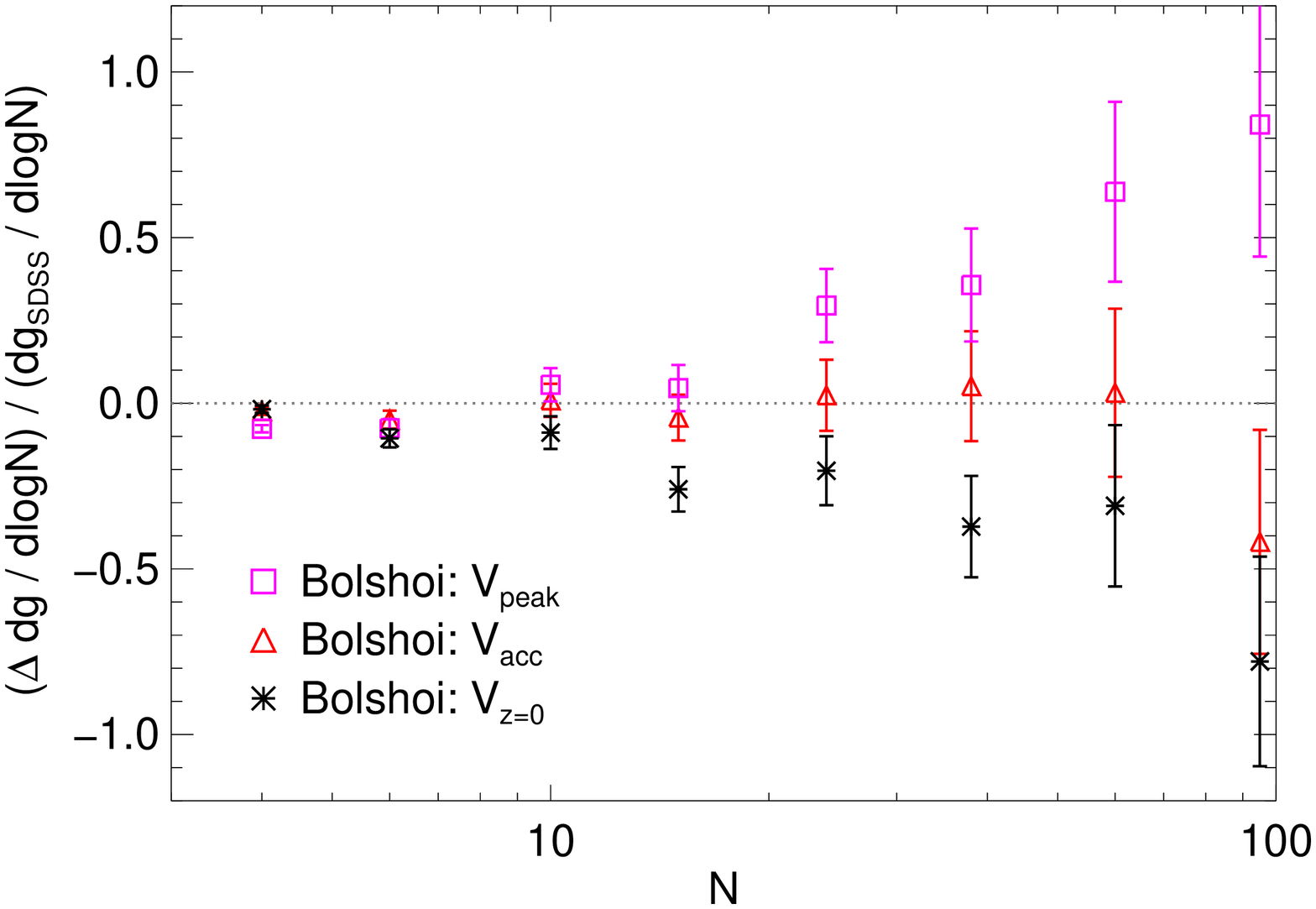}
\caption{
Group abundance as a function of the number of group members. In the {\em top} panel we illustrate a comparison of the differential group multiplicity function, $\dd g(N)/\dd\log N$ seen in the Mr19 SDSS catalog (blue diamonds) and that in our fiducial mock catalog (red triangles). Our fiducial mock was made with abundance matching on $\vacc.$ We also display results when the abundance matching is done on $\vpeak$ (top dashed curve) as well as $\vzero$ (bottom dashed curve). Error bars for these alternate SHAM models have been omitted as they are very similar to those in our fiducial model. In the {\em bottom} panel we plot the fractional difference between the predicted and observed $\dd g(N)/\dd\log N$ for each of these three models. All models depicted in this figure pertain to our fiducial scatter model, though we find no qualitative change to these results when exploring our alternate scatter model, or SHAM models without scatter.
}
\label{fig:gn}
\end{figure*}
%-----------------------------------------------------------------------------------------------------

In this section, we compare the predictions of abundance matching 
for the average number density of groups as a function of the 
number of group members, $N$.  We refer to this abundance 
as the group {\em multiplicity function} 
and represent it as $g(N)$. The differential multiplicity function, $\dd g(N)/\dd\log N,$ of the
the observed SDSS Mr19 sample is plotted in blue diamonds in the 
top panel of Fig.~\ref{fig:gn}.  We use nine bins evenly spaced in $\log N$ over the range $3\le N\le100.$ In all but the smallest richness bin, the observed multiplicity function 
is consistent with a power-law of $g(N) \propto N^{-2.5}$.

The errors in Fig.~\ref{fig:gn} were computed by bootstrap 
resampling of the group catalogs. For all of the results in this paper, 
our bootstrap errors are computed as the dispersion over $10^4$ bootstrap 
realizations of our group sample, where each realization is constructed by randomly 
selecting\footnote{We bootstrap resample {\em with replacement}, that is, we allow for the possibility of repeated draws of the same object.} $N_{\mathrm{s}}$ groups from the original sample, where $N_{\mathrm{s}}$ is the number of groups in the original sample. An alternative error estimation technique to bootstrapping is jackknifing, in which subsamples are drawn from specific sub-volumes of the simulation or survey. In jackknifing, the variations between the sample realizations are more closely tied to large-scale spatial variations than in the bootstrapped realizations, and so jackknifing is commonly used for statistics in which cosmic variance is thought to be the dominant source of error. We explored our error estimation using both jackknifing and bootstrapping, and found that for the group multiplicity function the errors are generally larger for the case of bootstrapping. This suggests that random errors in our group-finding algorithm due to, e.g., redshift-space projection effects dominate the variance in our sample. Regardless, the larger errors in the bootstrapping estimation method make this the more conservative choice between the two techniques.

In addition to depicting the SDSS Mr19 group multiplicity function, 
Fig.~\ref{fig:gn} shows how these data may be used to scrutinize 
empirical methods to assign galaxies to halos, such as SHAM. 
To illustrate this, we plot with red triangles the differential multiplicity 
function prediction of our fiducial SHAMacc mock catalog. The upper (lower) 
dashed curve in the top panel of Fig.~\ref{fig:gn} shows the SHAMpeak 
(SHAM0) prediction for the multiplicity function; error bars for these 
two models have been omitted as they are similar to those from our 
fiducial SHAMacc model. To further facilitate the illustration of the 
potential of $g(N)$ measurements to discriminate between different 
abundance matching prescriptions, in the bottom panel of 
Fig.~\ref{fig:gn} we plot the fractional difference between 
the predicted and observed differential multiplicity functions. All the points 
appearing in Fig.~\ref{fig:gn} trace models with our 
fiducial amount of scatter between $M_r$ and $\vmax.$

Our fiducial SHAMacc model is consistent with the observed multiplicity 
function (the difference between the two distributions is less than $1.5\sigma$).\footnote{Here and throughout, when we quote the statistical significance of a difference between two distributions, $x_{\mathrm{i}}$ and $y_{\mathrm{i}}, i=1,\dots,N_{\mathrm{dof}},$ we refer to the results of the following exercise. We measure the difference between the distributions in each bin, $d_{\mathrm{i}}\equiv(x_{\mathrm{i}}-y_{\mathrm{i}}),$ and estimate the errors as $\sigma(d_{\mathrm{i}})=\sqrt{\sigma(x_{\mathrm{i}})^{2} + \sigma(y_{\mathrm{i}})^{2}}.$ With $\Delta\chi^{2}=\Sigma_{i}[d_{\mathrm{i}}/\sigma(d_{\mathrm{i}})]^{2},$ we estimate the statistical significance of the difference between $x$ and $y$ as $P(\Delta\chi^2,N_{\mathrm{dof}}),$ where $P(x,N)$ is a $\chi^2$ distribution with $N$ degrees of freedom.}
On the other hand, the SHAM0 assignment with $\vsub=\vzero$ is a significantly poorer description of the 
data.  SHAM0 underestimates the abundances of all groups with $N \geq 5$.  
The significance of the difference between the SDSS group data and the predicted 
SHAM0 multiplicity function is greater than $5\sigma.$
These discrepancies hold true at similar levels in the alternate 
scatter models we studied.

This discrepancy may not be surprising 
because SHAM assignments based upon $\vzero$ have 
already been shown to be less effective at describing independent 
data than SHAM assignments based upon $\vacc$ 
\cite[e.g.][]{conroy_etal06,berrier_etal11,watson_etal11,reddick_etal12}.  
More interestingly, our SHAMpeak assignments of galaxies to halos with $\vsub=\vpeak$ 
do not describe the data as well as SHAMacc either.  
SHAMpeak models significantly {\em overestimate} the abundances of rich groups.
The statistical significance of this discrepancy is greater than $4\sigma.$ 

These results are interesting as a demonstration that the multiplicity functions 
of groups can be used as valuable statistics with which to constrain the connection 
between dark matter and galaxies. We can understand why $g(N)$ is a useful statistic to discriminate between SHAM mocks based on different halo properties as follows. Subhalos in the SHAMpeak model will have larger $\vmax$ values than subhalos in SHAMacc models, which in turn will have larger $\vmax$ values than those in SHAM0 models. Subhalos with larger $\vmax$ values  will be assigned brighter luminosities, implying that there will be more satellite galaxies in SHAMpeak (SHAMacc) models relative to SHAMacc (SHAM0) models, boosting the richness $N$ in host halos. This increase must come at the expense of a decrease in field ($N=1$) galaxies, and, indeed, we find that there are more field galaxies in SHAM0 models than in SHAMacc models, and more SHAMacc field galaxies than in SHAMpeak models.

The group multiplicity function test illustrated in Fig.~\ref{fig:gn} clearly 
indicates that the SHAMacc assignment with $\vsub=\vacc$ is consistent with SDSS 
group data while the alternative SHAM0 and SHAMpeak assignments cannot describe 
the SDSS group data adequately.  In particular, note that the SHAMacc results are 
straddled by the SHAMpeak and SHAM0 results, suggesting that it may be possible 
to use $g(N)$ measurements to constrain models of the mass stripped from satellite 
galaxies \cite[e.g.][]{watson_etal11}. The success of our SHAM models with $\vsub=\vacc$ 
motivates the choice of SHAMacc as our fiducial model.  

We have also investigated the role that scatter plays in the abundance matching 
prediction for $g(N).$ Each of the models depicted in Fig.~\ref{fig:gn} pertain 
to our fiducial model of scatter, which has $0.2$dex of scatter at the faint end 
and $0.15$dex at the bright end. However, we find no qualitative change to our results 
when using our alternate scatter model (which has a constant $0.1$dex of scatter) 
or models with no scatter. In particular, we find that for all the scatter models 
we explored, SHAMpeak models significantly overestimate the abundances of rich groups 
and SHAM0 models significantly underestimate the abundances of rich groups. 
Accordingly, we do not explicitly show the results for the alternative scatter models 
for the sake of brevity.  

In this subsection {\em only}, we study an alternate SDSS catalog that is 
complete to $\mr\leq-20,$ spans the redshift range $0.02<z<0.106,$ and has an 
effective volume of $\veff \simeq 2.3\times10^{7}$ $\mathrm{Mpc}^{3}/h^3$ to 
test the sensitivity of our conclusions to the sample selection. 
We constructed mock catalogs for this Mr20 catalog in exactly the same fashion as 
for Mr19, except that we imposed a brightness cut of $\mr\leq-20$ for the mock galaxies. 
Our findings for the Mr20 catalog are the same as for Mr19: 
the observed $g(N)$ is well-fit by a power law with exponent $\sim-2.5$ in all but 
the smallest richness bin; 
the multiplicity function predicted by our fiducial model of the Mr20 groups 
exhibits less than $1\sigma$ discrepancy with the data; 
SHAMpeak (SHAM0) significantly over-(under-)predicts the abundances of rich groups, 
with differences between these models and the data being greater than $4.5\sigma$ in each case.

The robustness of our conclusion that SHAMpeak models over-predict the abundance of rich groups is particularly interesting in light of recent results \citep{reddick_etal12} demonstrating that SHAMpeak models provide a much more accurate prediction for projected galaxy clustering on small scales relative to SHAMacc models. In fact, the authors in \citet{reddick_etal12} specifically trace the difference in the clustering predictions of these two models to the lower abundance of satellite galaxies in SHAMacc models. Yet, it is precisely this behavior that results in SHAMacc models providing more accurate predictions for the group multiplicity function in the regime $N\gtrsim20.$ These two results thus seem to be puzzlingly at odds with one another.

One possible explanation for the apparent discrepancy between our multiplicity function results and the clustering results in \citet{reddick_etal12} is that group identification is sensitive to observational systematics that we may be inadequately accounting for. The two most likely candidates for such systematics are fiber collisions and effects due to the survey edges in the SDSS galaxy sample that do not influence our mock galaxy sample. We have estimated the influence of both of these systematics on $g(N)$ and discuss our findings in detail in Appendix C. Briefly, we find that neither fiber collisions nor edge effects have nearly a significant enough impact on the multiplicity function to bring the SHAMpeak prediction for $g(N)$ in line with observations.

An additional possible explanation is that, once the effect of orphan galaxies are included in the construction of the mock catalogs, the clustering prediction of SHAMacc models is brought into better agreement with the SDSS measurements of the two-point function. We provide further discussion of this caveat in \S~\ref{subsection:orphans}.

\begin{figure}
\centering
\includegraphics[width=8.0cm]{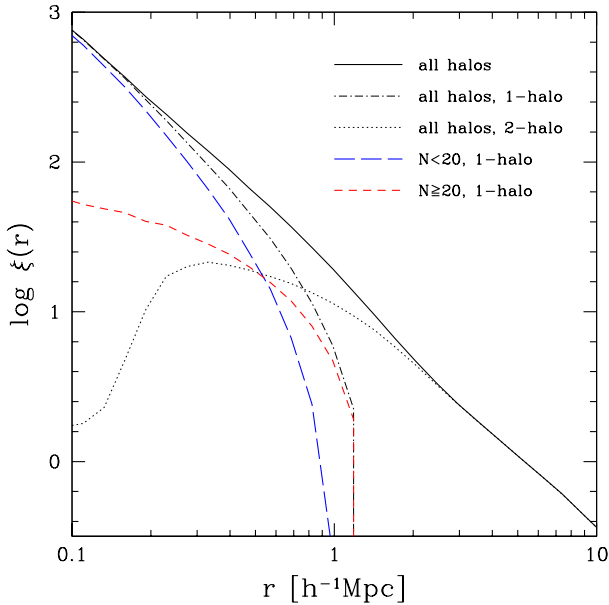}
\caption{
The correlation function of mock galaxies in the Bolshoi simulation using our fiducial SHAMacc abundance matching model. The solid black line shows the correlation function of all mock galaxies, while the thinner dot-dashed and dotted black lines show the 1-halo (pairs of galaxies residing in the same halo) and 2-halo (pairs of galaxies residing in separate halos) contributions to the correlation function, respectively. The blue long-dashed and red short-dashed lines break the 1-halo term into contributions from galaxies that live in halos with $N<20$ and $N\geq20$ galaxies, respectively. The figure shows that galaxies in $N\geq20$ halos never dominate the correlation function.
%The correlation function of mock galaxies in the Bolshoi simulation using our fiducial SHAMacc abundance matching model. The solid black line shows the correlation function of all mock galaxies, while the thinner dot-dashed and dotted black lines show the 1-halo (pairs of galaxies residing in the same halo) and 2-halo (pairs of galaxies residing in separate halos) contributions to the correlation function, respectively. The blue long-dashed and red short-dashed lines break the 1-halo term into contributions from galaxies that live in halos with $N<20$ and $N\geq20$ galaxies, respectively. The figure shows that galaxies in $N\geq20$ halos never dominate the correlation function. At small scales ($r\lesssim0.5\hmpc$), the correlation function is dominated by 1-halo pairs of galaxies in $N<20$ halos, while at larger scales the correlation function is dominated by 2-halo pairs. This behavior is particularly interesting since the multiplicity function prediction of SHAMpeak models only begins to deviate significantly from the observed $g(N)$ in groups with $N\gtrsim20$ members.
}
\label{fig:xin}
\end{figure}

We propose the following reconciliation between our results and those in \citet{reddick_etal12}: {\em two-point galaxy clustering is insensitive to dark matter halos that host $N\gtrsim20$ galaxies.} We support this claim in Fig.~\ref{fig:xin}, in which we study the contribution to the two-point correlation function $\xi(r)$ from samples of halos in Bolshoi with different richness threshold cuts. For our fiducial SHAMacc mock catalog, we plot the three-dimensional real-space correlation function $\xi(r)$ (solid black line). We also plot the 1-halo term, which only counts pairs of galaxies that reside in the same halo (dot-dashed black line), and the 2-halo term, which only counts pairs of galaxies that reside in distinct halos (dotted black line). Finally, we show the separate contributions to the 1-halo term of galaxies that live in $N<20$ halos (long-dashed blue line) and $N\geq20$ halos (short-dashed red line). At scales smaller than $\sim0.5\hmpc,$ the correlation function is dominated by 1-halo pairs of galaxies in small ($N<20$) halos, while at larger scales, the correlation function is dominated by the 2-halo term. 1-halo pairs in large ($N\geq20$) halos only contribute significantly to $\xi(r)$ in a narrow range of scales, and even in this regime they are subdominant.

The failure of SHAMpeak models at predicting the group multiplicity function in the $N\gtrsim20$ regime is not at odds with the \citet{reddick_etal12} finding that SHAMpeak provides the most accurate prediction for $\xi(r)$ on small-scales, because this failure only occurs in halos that have very little influence on the correlation function. However, combining the \citet{reddick_etal12} result that SHAMacc models do not adequately describe two-point clustering with our result that SHAMpeak models do not adequately describe the group multiplicity function implies that {\em no previously explored SHAM model successfully predicts both $\xi(r)$ and $g(N).$} In \S~\ref{section:discussion} we discuss the implications of this finding, as well as an important caveat concerning orphan galaxies.

%---------------------------------------------------------------------------------------------------
%-----------------------------------------------------------------------------------------------------

%---------------------------
\subsection{Field \& Group Galaxy Luminosity Functions}
\label{subsection:groupfield}
%---------------------------

The SHAM method for assigning galaxies to halos results in mock luminosity 
functions that match observed luminosity functions by construction.  However, 
this procedure does not guarantee agreement between luminosity functions 
that are conditioned on a specific galaxy property or environment.  In this 
section, we consider a simple distinction in galaxy environment based directly 
on our group catalogs.  Specifically, we explore the luminosity function of galaxies 
conditioned upon whether or not the galaxy is identified as a member of a group. 
We refer to the luminosity function constructed from all galaxies residing in 
groups as $\Phi_{\mathrm{group}}(L)$. 
For group galaxies, we explore several different richness threshold cuts defining group membership; where applicable, we will explicitly state the particular richness threshold used in the definition of group galaxies.
Likewise, we refer to all galaxies that we do not identify as members of a 
group as ``field'' galaxies; to be explicit, thoughout this paper ``field" galaxies are defined as those galaxies found in groups with only a single member. 
 We denote the luminosity function conditioned on the 
galaxy being a member of the field as $\Phi_{\mathrm{field}}(L)$.  
SHAM predictions for $\Phi_{\mathrm{group}}(L)$ and $\Phi_{\mathrm{field}}(L)$ are 
{\em not} guaranteed to match observational determinations of these quantities, so 
this is a test of the allocation of galaxies to group and field environments 
by SHAM methods.

\begin{figure*}
\centering
\includegraphics[width=8.0cm]{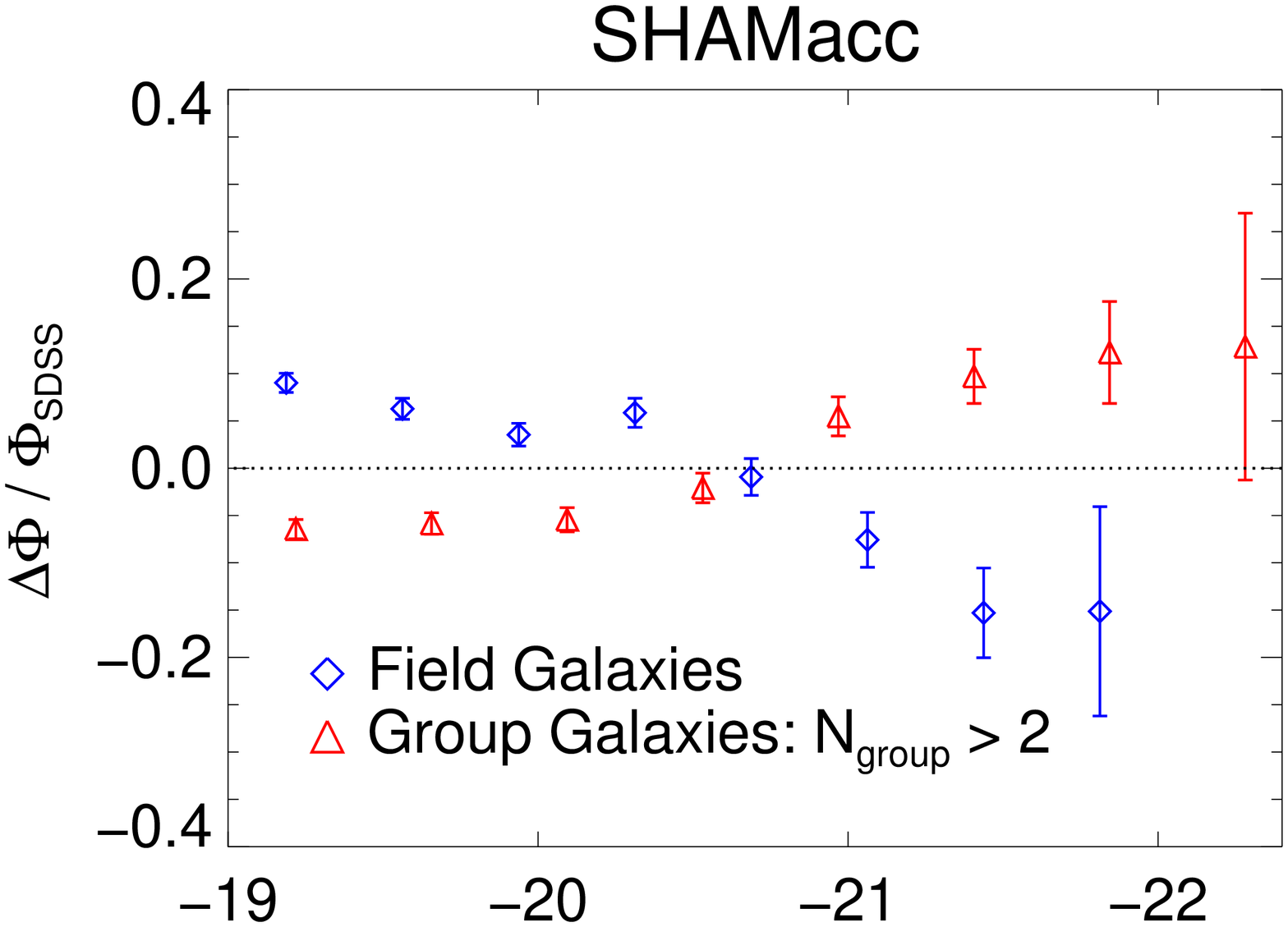}
\includegraphics[width=8.0cm]{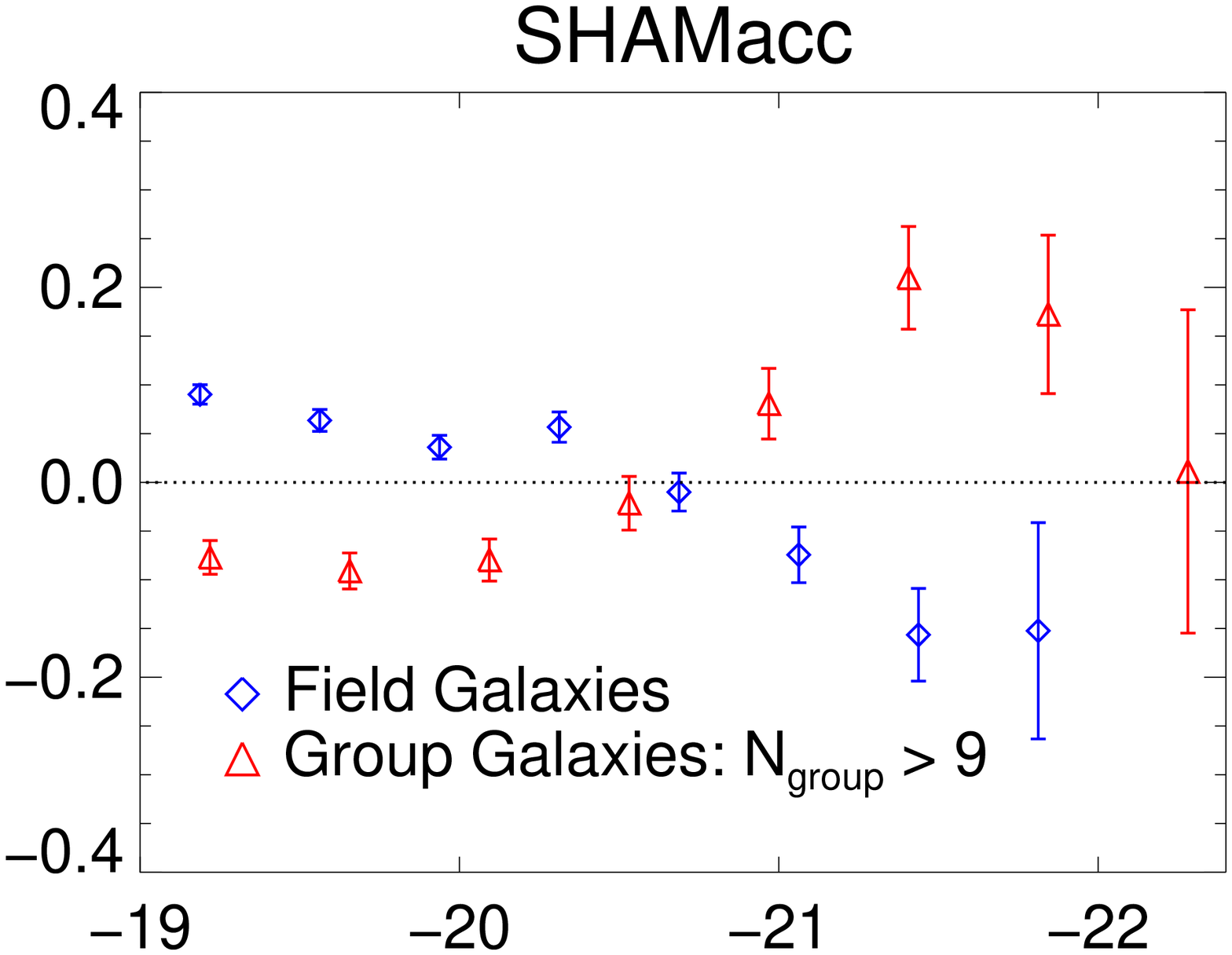}
\includegraphics[width=8.0cm]{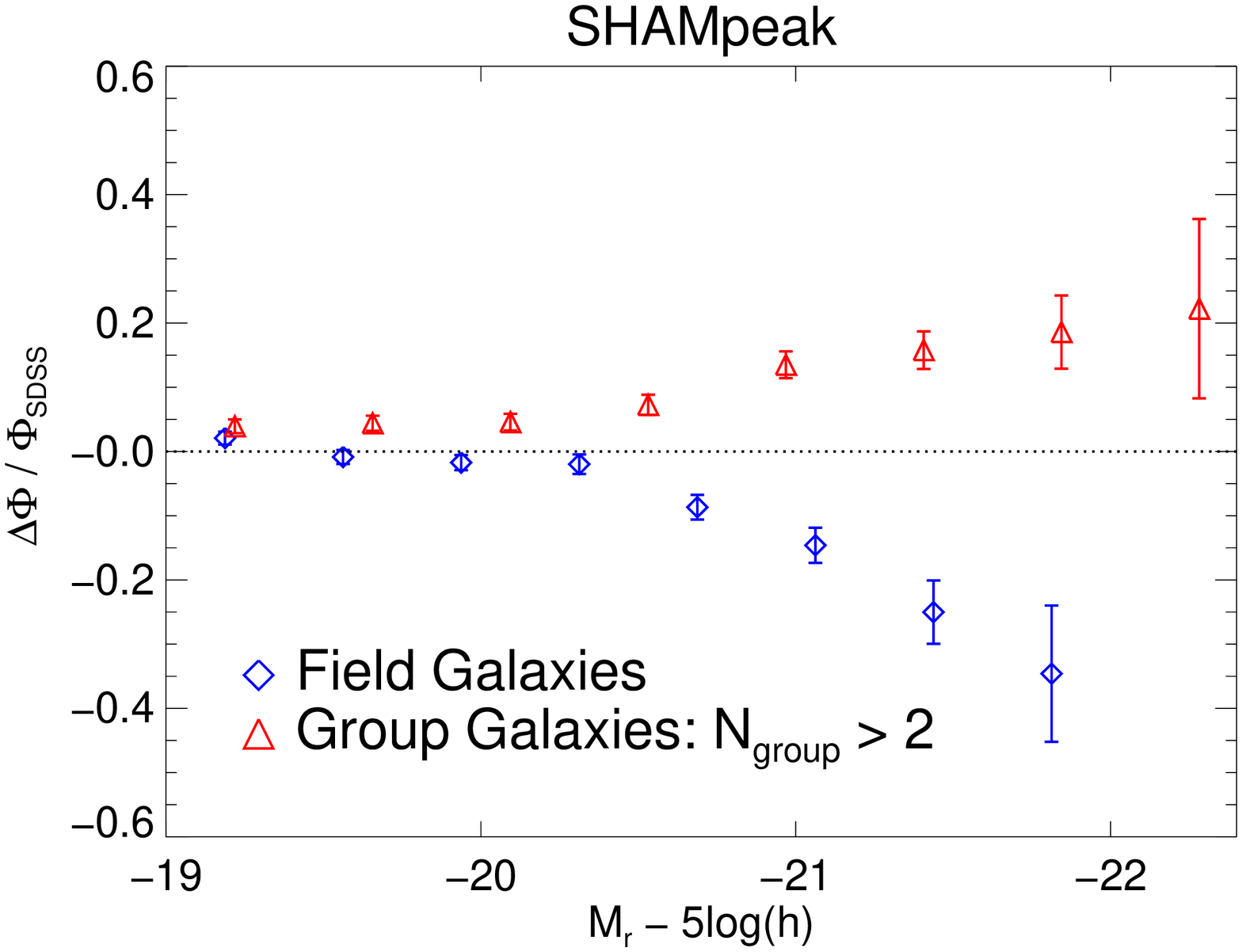}
\includegraphics[width=8.0cm]{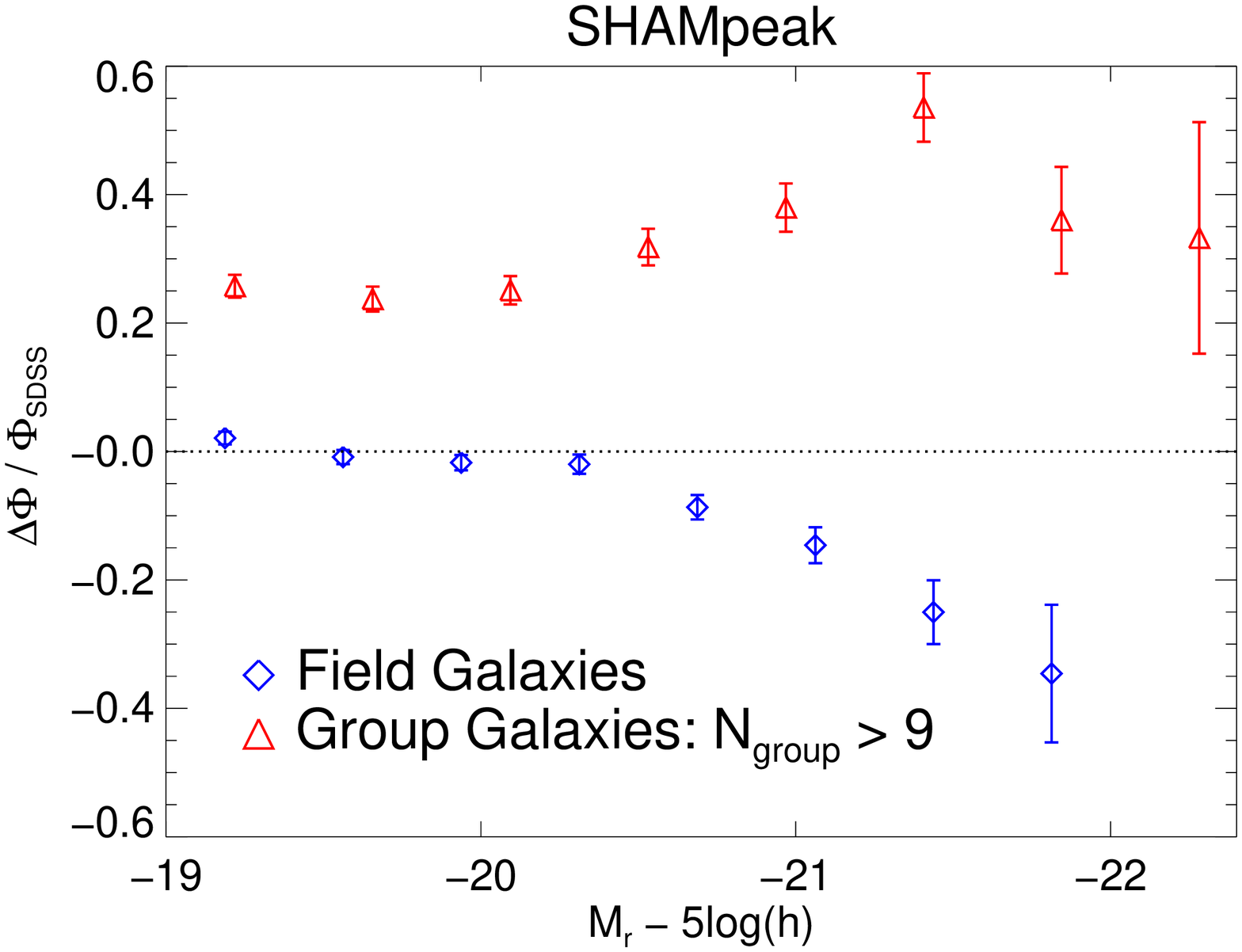}
\caption{
Fractional differences between the field and group galaxy
luminosity functions predicted by abundance matching and observed 
in SDSS. The sign convention adopted in the vertical axes is defined by $\Delta\Phi\equiv \Phi_{\mathrm{mock}}-\Phi_{\mathrm{SDSS}}.$
%The fractional differences for field (group) galaxies appear as blue diamonds (red triangles). For this purpose, field galaxies are defined to be isolated galaxies that reside in groups with $N=1$ members, and group galaxies are defined by the richness cut given in the legend of each panel. 
Results pertaining to our fiducial SHAM model, SHAMacc, based on abundance matching with $\vsub=\vacc$, appear in the top panels. Results pertaining to SHAM using $\vpeak$, SHAMpeak, appear in the bottom panels. 
%The left panels show results for groups with multiplicity $N_{group}>2$ while the right panels show results for $N_{group}>9$. 
The group (field) galaxy luminosity functions in all of our SHAM mocks is systematically too bright (dim). No SHAM model predicts either the group or field galaxy luminosity function to better than $\sim10\%$ at almost any luminosity.
 }
\label{fig:groupfield}
\end{figure*}

Figure~\ref{fig:groupfield} shows comparisons between our predicted, SHAM luminosity 
functions for group galaxies and field galaxies and the corresponding 
observed luminosity functions.  To highlight differences, the quantity 
shown on the vertical axis of all panels in Fig.~\ref{fig:groupfield} is the 
fractional difference $\Delta\Phi/\Phi_{\mathrm{SDSS}}$ between the predictions of the 
SHAM mocks and the observations, 
where $\Delta\Phi(L) \equiv \Phi_{\mathrm{mock}}(L)-\Phi_{\mathrm{SDSS}}(L),$ 
so that points in Fig.~\ref{fig:groupfield} with positive vertical axis values 
correspond to luminosities where SHAM over-predicts the abundance of galaxies of that 
brightness, and conversely for negative values of $\Delta\Phi(L)$.  
Blue diamonds show $\Delta\Phi/\Phi_{\mathrm{SDSS}}$ for field galaxies, red triangles show the same 
for group galaxies. Group galaxies defined by a richness cut of 
$N_{group} \geq 3$ appear in the left columns, $N_{group}\geq 10$ appear in the 
right columns. Our fiducial model, SHAMacc, which abundance matches on $\vacc$, 
appears in the top panels; the alternate, SHAMpeak model, which uses $\vsub=\vpeak$, 
appears in the bottom panels; all abundance matching prescriptions in 
Figure~\ref{fig:groupfield} were constructed with our fiducial scatter. 
The error bars on $\Phi_{\mathrm{SDSS}}(L)$ and $\Phi_{\mathrm{mock}}(L)$ have each been 
estimated by bootstrap resampling as in \S~\ref{subsection:gn}. 
We reiterate that the differences shown in this plot are strictly due to 
the separation of field galaxies from group galaxies because our SHAM 
implementation {\em guarantees} that the overall luminosity function matches 
the data {\em exactly}.\footnote{Since the overall $\Phi(L)$ is exactly correct in our mocks, the reader may wonder why the sum $\Phi_{\mathrm{group}}(L)+\Phi_{\mathrm{field}}(L)$ appears to be inconsistent with zero in Fig.~\ref{fig:groupfield}. The reason is due to our definition of group galaxies: in the left panels, galaxies in groups with richness $N=2$ are not included in $\Phi_{\mathrm{group}};$ in the right panels, galaxies in $N=2,\dots,9$ groups are not included.  When $\Phi_{\mathrm{group}}(L)$ is defined so that galaxies in groups of all richnesses $N\geq2$ are included, the sum $\Phi_{\mathrm{group}}(L)+\Phi_{\mathrm{field}}(L)$ is, indeed, consistent with zero.}

As Fig.~\ref{fig:groupfield} shows, our fiducial SHAMacc model, 
the baseline SHAM procedure which provided an accurate description of the group multiplicity function in \S~\ref{subsection:gn}, 
systematically {\em overestimates} the abundance of dim, field galaxies and {\em overestimates} 
the abundance of bright, group galaxies. 
At virtually all luminosities, SHAMacc models fail to predict both $\Phi_{group}(L)$ and $\Phi_{field}(L)$ at the level of $10-20\%;$ the biggest failures occur in SHAMpeak models, which fail at the $40\%$ level in large groups.
Notice that this qualitative 
conclusion holds irrespective of the choice for the minimum number of member 
galaxies required in order to have the group members be included in the 
group luminosity function. In fact, the discrepancies grow larger as the number 
of group members grows, emphasizing that the SHAM excesses of bright, group 
galaxies grow more egregious for larger groups (particularly for SHAMpeak).  
These results highlight at least one weakness of 
the SHAM procedure for exploring the connection between galaxies and dark matter halos.  
The widely-used SHAM procedure that adequately describes 
low-redshift galaxy clustering {\em does not allocate luminosities 
to group and field halos in a manner consistent with observations}. 

%---------------------------------------------------------------------------------------------------
%-----------------------------------------------------------------------------------------------------

\begin{figure*}
\centering
\includegraphics[width=8.0cm]{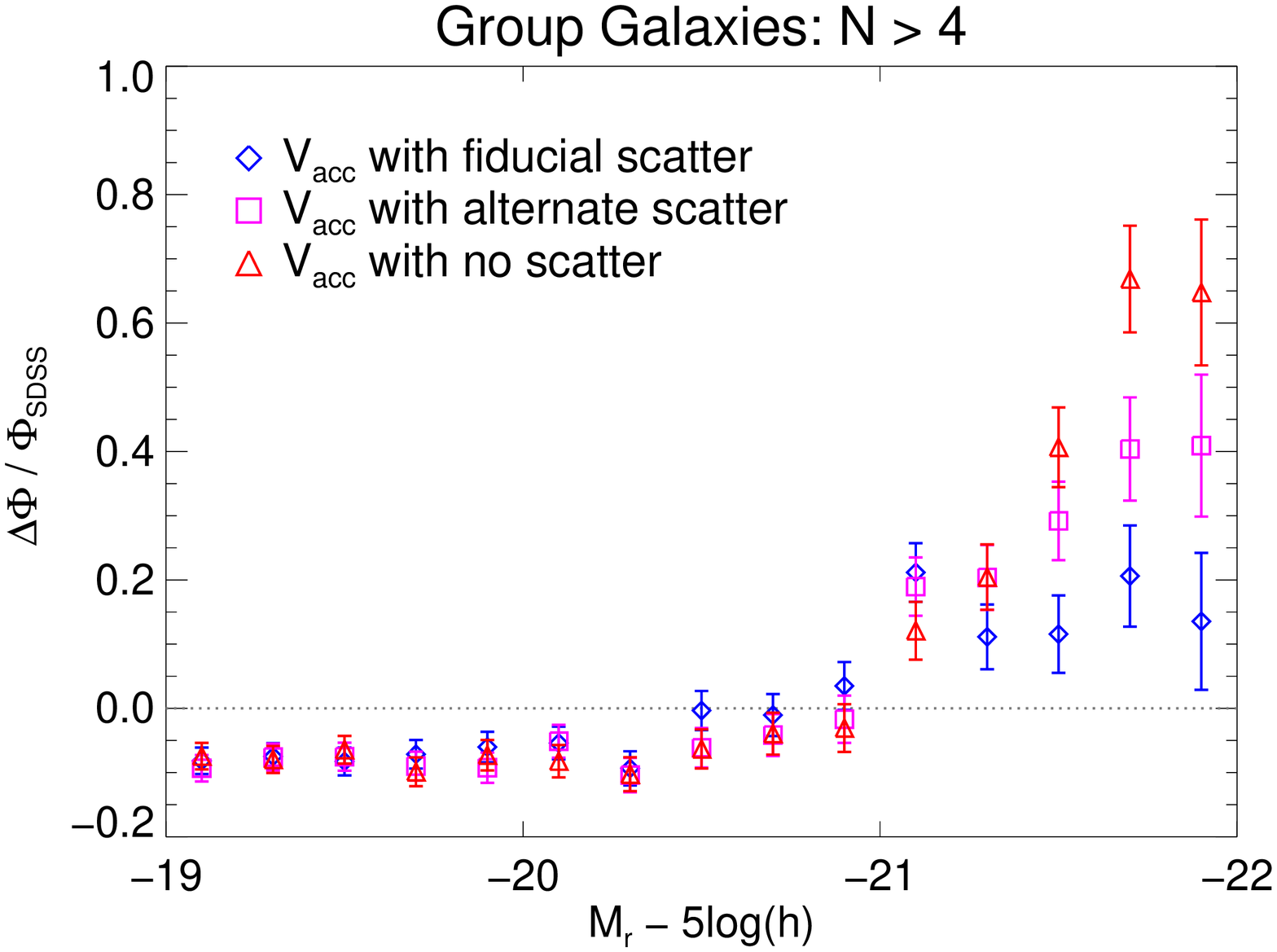}
\includegraphics[width=8.0cm]{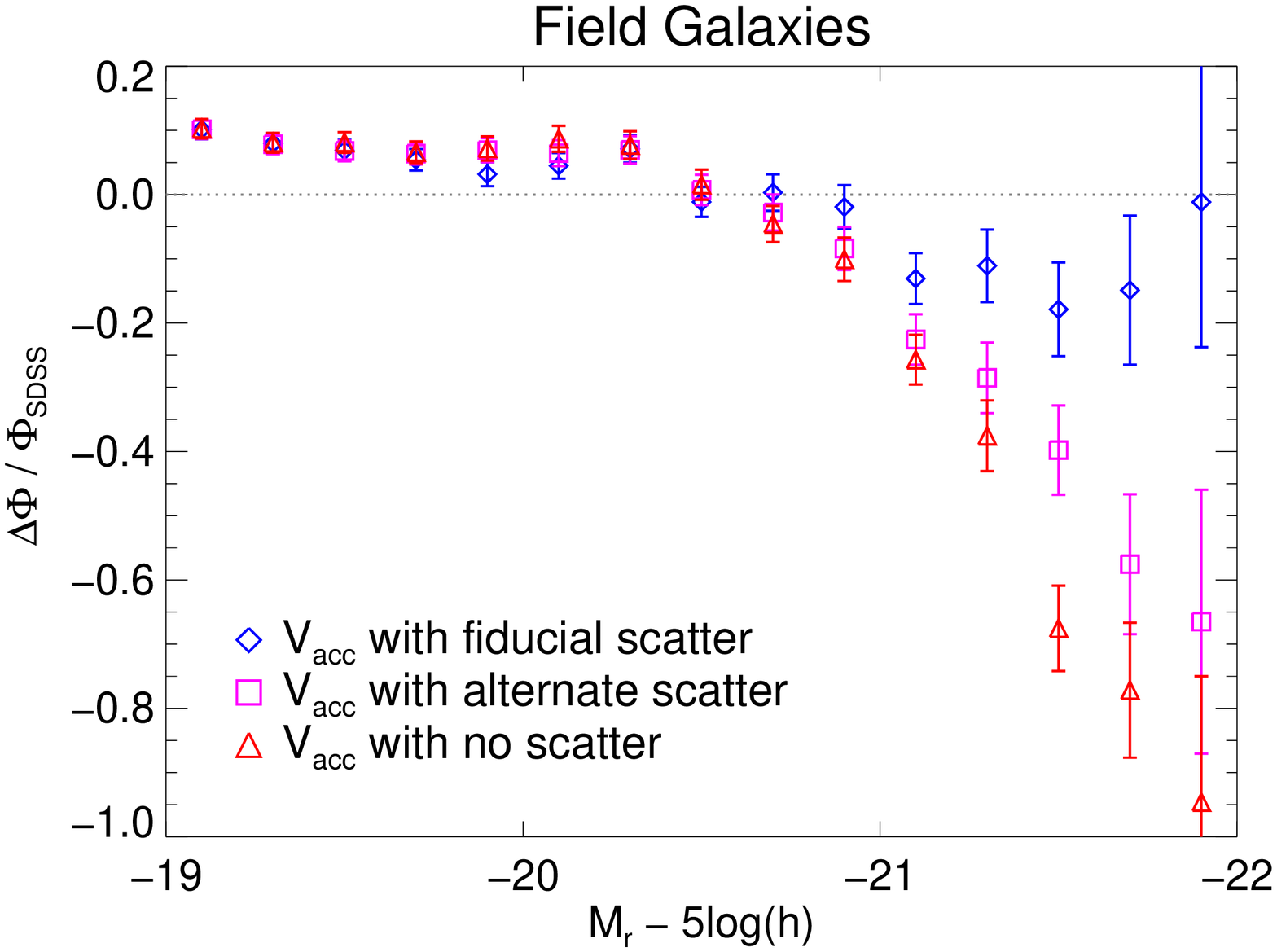}
\caption{
Fractional difference between the group environment-conditioned luminosity functions seen in 
SDSS and the predictions for SHAMacc with our fiducial scatter model (blue diamonds), 
our alternate scatter model (magenta squares), and no scatter (red triangles). 
The sign convention adopted in the vertical axes is defined by $\Delta\Phi\equiv \Phi_{\mathrm{mock}}-\Phi_{\mathrm{SDSS}}.$
%For this purpose, field galaxies are defined to be isolated galaxies that reside in groups with $N=1$ members,  and the group galaxy sample is defined by requiring that each galaxy in the sample reside in a group with $N>4$ members. 
This figure is similar to Fig.~\ref{fig:groupfield}, but here we address the influence of scatter in the SHAMacc predictions for the group and field luminosity functions.  
Regardless of the amount of scatter, errors at levels of $\gtrsim10\%$ persist in all SHAM model predictions for 
$\Phi_{\mathrm{field}}(L)$  and $\Phi_{\mathrm{group}}(L).$ 
}
\label{fig:groupfieldscatter}
\end{figure*}

We also explore the influence that scatter between absolute magnitude and halo circular velocity has on 
the SHAM predictions for the group and field luminosity functions. In Figure~\ref{fig:groupfieldscatter}, 
we again plot the fractional difference between the observed and predicted $\Phi_{\mathrm{group}}(L)$ (left panel) 
and $\Phi_{\mathrm{field}}(L)$ (right panel), this time comparing results between mocks made with 
different amounts of scatter. The SHAMacc abundance matching procedure was used for all models 
depicted in Figure~\ref{fig:groupfieldscatter}. Blue diamonds in Fig.~\ref{fig:groupfieldscatter} 
give results pertaining to our fiducial scatter model, 
which has $0.2$dex of scatter at the faint end and $0.15$dex of scatter at the bright end, 
magenta squares designate our alternate model with a constant $0.1$dex of scatter, 
and red triangles depict our SHAMacc mock without scatter. 
Group galaxies in Fig.~\ref{fig:groupfieldscatter} have been 
defined by the richness cut $N_{group}>4.$ 

%---------------------------------------------------------------------------------------------------
%-----------------------------------------------------------------------------------------------------

Evidently, the amount of scatter in SHAM significantly influences the predictions 
for $\Phi_{\mathrm{group}}(L)$ and $\Phi_{\mathrm{field}}(L)$ at the bright end, 
but has little role in the group and field luminosity functions at 
the faint end. A more exhaustive exploration of different prescriptions for 
the scatter may yield a model that correctly predicts the abundance of {\em bright} 
group and field galaxies. Any such model would require significantly larger scatter 
than the fiducial model we have used in this work, which is, in turn, based on the 
success of \citet{trujillo-gomez_etal11} in describing galaxy clustering and 
a variety of other properties using SHAM. 
However, the robustness of our results to differences at the faint end of the 
luminosity functions indicates that this is a generic weakness of SHAM that cannot 
be overcome by adding scatter alone. We discuss this point further in \S~\ref{section:discussion}, as well as the role that orphan galaxies may play in mitigating this discrepancy.

%---------------------------
\subsection{Magnitude Gap Statistics}
\label{subsection:pgap}
%---------------------------

In \S~\ref{subsection:groupfield}, we compared the luminosity functions 
of group galaxies predicted by SHAM to that exhibited in SDSS group data.  In this section, we 
explore a group statistic that is related to the group luminosity function 
known as the ``magnitude gap.''  We define the magnitude gap to be 
the difference in r-band absolute magnitude between the two brightest non-fiber collided 
members of any group, $\monetwo = M_{r,2}-M_{r,1}$, where 
$M_{r,\mathrm{i}}$ is the r-band absolute magnitude of the $\mathrm{i}^{th}$ brightest non-fiber collided 
group member. In particular, we will be interested in $\Phi(\monetwo),$ the 
statistical distribution of magnitude gaps, defined so that $\Phi(\monetwo)\dd\monetwo$ 
represents the number density of galaxy groups with magnitude gap $\monetwo$ in a bin of width $\dd\monetwo.$

As we discussed in \S~\ref{section:introduction}, the magnitude gap abundance 
statistic has received significant attention in recent literature.  
Of course, it is possible to enumerate the 
absolute magnitudes of all group members, but the magnitude 
gap has received particular attention for a number of reasons.  
These include:  
(1) the simplicity of a single statistic; 
(2) dynamical friction timescales 
vary in inverse proportion to galaxy mass, so the largest satellites 
merge the most quickly, making magnitude gap a possible indicator 
of the dynamical age of a group; 
and (3) the magnitude gap may help 
to refine mass estimates for optically-identified clusters \citep{hearin_etal12,more12}.  

We will be particularly interested in testing the success of SHAM models at predicting $\Phi(\monetwo),$ and so we begin our investigation in \ref{subsubsection:bcglf} by first studying the the luminosity function of the brightest and next-brightest group galaxies, as it is these galaxies from which the magnitude gap of a group is computed. We discuss the natural, statistical correlation between $\monetwo$ and group richness in \ref{subsubsection:gaprich}, motivating the multiplicity function-matching technique that we use throughout the rest of this section. We present results for the predicted and observed $\Phi(\monetwo)$ in \ref{subsubsection:phigap}, and focus on the large-gap tail of this distribution in \ref{subsubsection:ffos}.

%---------------------------
\subsubsection{Luminosity Function of Brightest and Next-Brightest Group Galaxies}
\label{subsubsection:bcglf}
%---------------------------

As discussed above, the magnitude gap of a group is defined as the difference in r-band magnitude between the group's two brightest (non-fiber collided) members. Thus in our study of magnitude gaps, the luminosity function of brightest and next-brightest group galaxies is of interest. To measure these functions, we proceed by first rank-ordering the members of each group by their brightness; for convenience, we denote the luminosity of the $\ith$ brightest group member as $L_i.$ After this rank-ordering, we proceed to measure the luminosity function of the brightest and next-brightest members of each group, denoted by $\Phi(L_1)$ and $\Phi(L_2),$ respectively. 

\begin{figure*}
\centering
\includegraphics[width=8.0cm]{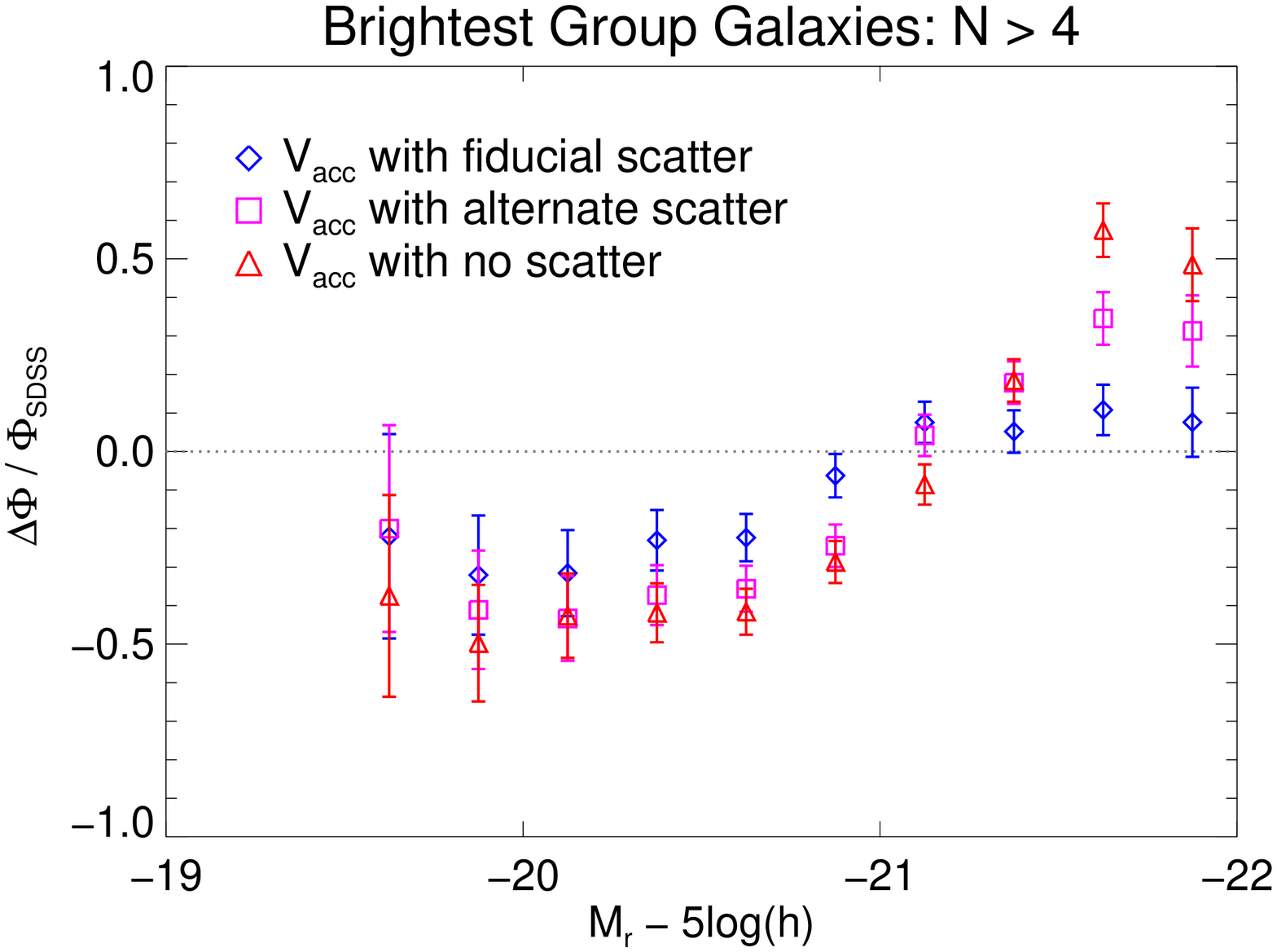}
\includegraphics[width=8.0cm]{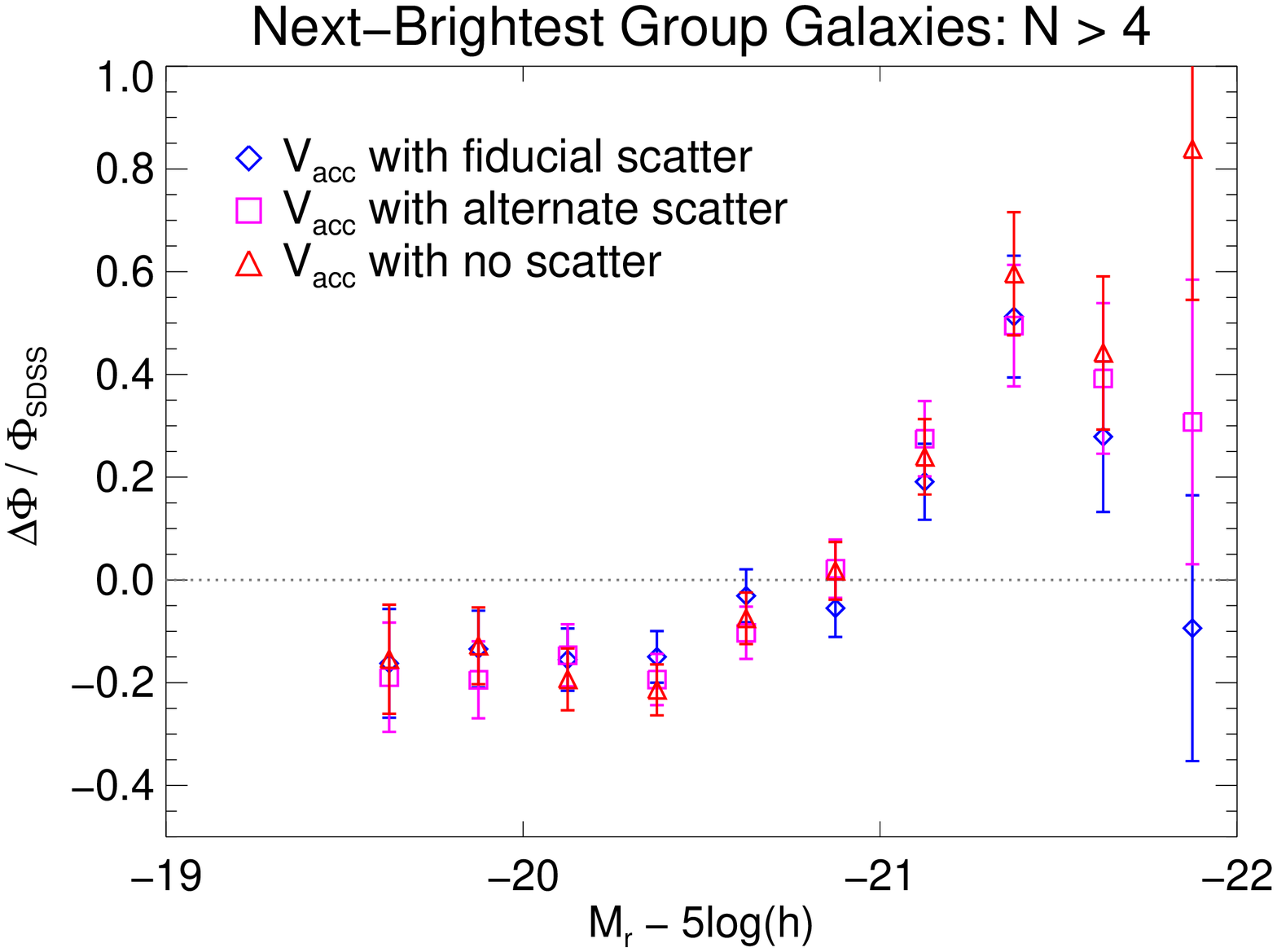}
\caption{
After rank-ordering the members of each group by their 
brightnesses, we have measured the luminosity function of the brightest groups members, $\Phi(L_1),$ as 
well as that of the next-brightest group members, $\Phi(L_2).$ 
In the left (right) panel we plot the fractional difference between the observed and predicted $\Phi(L_1)$ ($\Phi(L_2)$) . The sign convention adopted in the vertical axes is defined by $\Delta\Phi\equiv \Phi_{\mathrm{mock}}-\Phi_{\mathrm{SDSS}},$ as in Fig.~\ref{fig:groupfield}.
The group galaxy sample is defined by requiring that each galaxy in the sample reside 
in a group with $N>4$ members. The luminosity function of the brightest and next-brightest group members predicted by SHAM models is systematically too bright, a trend that is qualitatively similar to the errors in the SHAM predictions for $\Phi_{group}(L),$ discussed in \S~\ref{subsection:groupfield}.
}
\label{fig:bcglf}
\end{figure*}

We illustrate our results in Figure \ref{fig:bcglf}, in which we show the fractional difference between the predicted and observed $\Phi(L_1)$ in the left panel and between the predicted and observed $\Phi(L_2)$ in the right panel. We restrict attention to the rank-ordered luminosity function of galaxies found in groups with $N>4$ members, though plots made from samples with different richness cuts are very similar. 
Qualitatively, the trends 
in both panels reflect the sense of the errors on $\Phi_{\mathrm{group}}(L)$ 
shown in the left panel of Fig.~\ref{fig:groupfieldscatter}. SHAM predicts brightest group galaxies and 
next-brightest group galaxies that are significantly too bright on average, 
compared to observations.  Moreover, this conclusion is insensitive to scatter, 
so it is, again, a {\em generic weakness of the SHAM assignments}.  

%---------------------------------------------------------------------------------------------------
%-----------------------------------------------------------------------------------------------------

We present the results in 
Fig.~\ref{fig:bcglf} in large part because our results in 
the following sections focus on $\Phi(\monetwo),$ the relative luminosity of the 
brightest and next-brightest group members, and so the errors illustrated 
in Fig.~\ref{fig:bcglf} are germane to all of our results pertaining to the 
SHAM prediction for $\Phi(\monetwo).$ Before proceeding, we explicitly note that $\Phi(L_1)$ and $\Phi(L_2)$ {\em do not determine a unique distribution of magnitude gaps} $\Phi(\monetwo).$ Galaxy luminosities 
can be partitioned among group members in an infinite variety of ways that 
all lead to the same average group luminosity function. The magnitude gap distribution $\Phi(\monetwo)$ is determined not only by the global distributions $\Phi(L_1)$ and $\Phi(L_2)$ exhibited by the group sample, but also by the correlation between $L_1$ and $L_2$ within the groups. Such correlations encode, for example, whether groups with a brighter-than-average $L_1$ tend also to have a brighter-than-average $L_2.$ Thus measurements of $\Phi(\monetwo)$ provide additional information about the imprint of group assembly on the luminosity of galaxies that is not contained in the distributions  $\Phi(L_1)$ and $\Phi(L_2).$ We will return to this point and expand upon it in \S~\ref{section:mcs}.

%---------------------------
\subsubsection{Magnitude Gap and Richness}
\label{subsubsection:gaprich}
%---------------------------

\begin{figure}
\centering
\includegraphics[width=8.0cm]{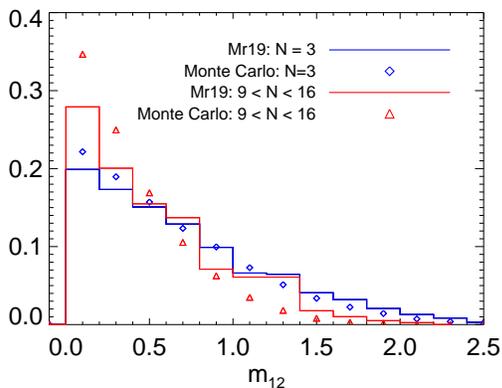}
\caption{
Trend of magnitude gap with richness. 
The blue (red) histogram traces the abundance of groups as 
a function of magnitude gap, $\Phi(\monetwo)$ (normalized to have unit area), exhibited by 
groups with richness $N=3$ ($9<N<16$). The blue diamonds and red triangles 
trace $\Phi(\monetwo)$ of the corresponding Monte Carlo randomizations of the groups, in which the brightness of group members are randomly drawn from a universal luminosity function. This figure demonstrates the basic trend of $\monetwo$ with $N,$ that  
richer groups tend to have smaller magnitude gaps, motivating the multiplicity function-matching methodology employed throughout the rest of this section.
}
\label{fig:gaprich}
\end{figure}

The abundance of groups by magnitude gap is a rapidly declining function of the gap, 
with approximately 90\% of all groups with $N \geq 3$ members in the 
Mr19 sample having a magnitude gap smaller than $\monetwo = 1.5.$ However, gap abundance 
depends sensitively on group richness, as demonstrated in
Figure~\ref{fig:gaprich}, where we show a histogram (normalized to have unit area) 
of $\Phi(\monetwo)$ exhibited by galaxy 
groups in two different richness ranges. 
The blue histogram traces $\Phi(\monetwo|N=3),$ the gap abundance of groups with $N=3$ members, 
while the red histogram traces $\Phi(\monetwo|9<N<16).$ A comparison of the two histograms 
provides a demonstration of the richness-dependence of $\monetwo$ abundance: richer groups 
tend to have smaller magnitude gaps. This trend has been demonstrated previously in the literature 
\cite[e.g.,][]{paranjape_sheth11,donghia_etal05,tavasoli_etal11}. 

%---------------------------------------------------------------------------------------------------
%-----------------------------------------------------------------------------------------------------

A simple way to gain insight into the sense of this trend is 
to consider a toy model universe in which galaxy groups are 
assembled by randomly drawing galaxy luminosities from a global 
luminosity function $\phiglo(L)$ (for example, a Schechter function).  As the richness $N$ of the toy groups 
increases, the number of random draws from $\phiglo$ increases, and the probability 
that a very bright member is drawn increases. Denoting the luminosity of the $\ith$ brightest 
members as $L_i,$ the expectation value of $L_i$ becomes brighter with increasing $N.$ As the 
number of random draws increases, the expectation value of $L_1$ is the first to become brighter 
than $L_*,$ the exponential cutoff of $\phiglo,$ and when this occurs there is a rapid decrease 
in the rate at which the expectation value of $L_1$ brightens with increasing $N.$  The 
reason for this rapid decrease is because of the exponential suppression in the luminosity 
function for galaxies with luminosties exceeding $L_*$; most draws correspond to lower luminosities.  
A relatively larger number of draws is required for the expectation value of $L_2$ 
to exceed $L_*,$ and so as $N$ continues to increase the expectation value of the 
ratio $L_1/L_2$ decreases.
%\footnote{Note that this intuitive explanation also demonstrates that the root cause of the shrinking of the magnitude gap with $N$ random draws is that the slope of $\phiglo$ steepens as the brightness increases. If $\phiglo$ did not have this property, for example if the global luminosity function were a simple power law, $\phiglo(L)\propto L^{\alpha},$ then the expectation value of the magnitude gap in the random draw universe would be entirely independent of $N$ since the expectation value of $L_1$ and $L_2$ would each brighten with increasing $N$ at the same rate.}

To illustrate this point explicitly, the blue diamonds and red triangles 
in Figure~\ref{fig:gaprich} show histograms of $\Phi(\monetwo)$ in Monte Carlo (MC) 
realizations of this toy universe. For each observed Mr19 group of richness $N$, 
we have constructed $1000$ realizations of the group by randomly drawing $N$ times 
from the {\em observed} $\Phi(L|N\geq3)$ (note that the exact, observed luminosity function of Mr19 galaxies 
is used and not a Schechter function approximation). 
Thus, for group samples in each richness bin plotted in Fig.~\ref{fig:gaprich}, the multiplicity functions of the observed 
and MC groups match {\em exactly}. The difference between the r-band magnitudes 
of the brightest two draws gives the $\monetwo$ value of the MC group. We then estimate $\Phi(\monetwo|N)$ of the random-draw universe simply by computing the mean abundance of the MC groups in bins of $\monetwo.$
Blue diamonds give the $\Phi(\monetwo|N=3)$ that results from this 
exercise, red triangles represent $\Phi(\monetwo|9<N<16).$ 

Of course, the toy model we illustrate with our Monte Carlo ignores any evolution of, or interactions between, 
group members during the process of group formation, and so it should not be surprising that the Monte Carlo predictions do not exactly trace the histograms in Figure~\ref{fig:gaprich}. However, the broad similarity 
between $\Phi(\monetwo|N)$ in the MC and in Mr19 data demonstrate 
that this model nonetheless provides a reasonable approximation of the relationship 
between richness and magnitude gap. 

\citet{paranjape_sheth11} emphasized the critical role that richness plays in any detailed study of $\monetwo,$ pointing out that $N$ is the appropriate group property to use to condition the group sample whose magnitude gap distribution is under consideration. We follow this methodology throughout paper. This distinguishes our work from previous measurements of the magnitude gap \cite[e.g.,][]{yang_etal08}, and is one of the salient features that make our measurements of gap abundance the most precise and complete in the literature to date.

%---------------------------
\subsubsection{Magnitude Gap Distribution}
\label{subsubsection:phigap}
%---------------------------

One of the chief goals in this paper is to demonstrate the utility of the 
observed gap abundance $\Phi(\monetwo)$ for constraining the galaxy-dark matter connection, 
with a particular focus on SHAM-based models. A basic consequence of the 
relationship between gap and richness is that the multiplicity function $g(N)$ 
plays a critical role in $\phi(\monetwo).$ 
As shown in \S~\ref{subsection:gn}, SHAMpeak and SHAM0 models 
over- and under-predict  $g(N)$ at moderate and large values of group richness, respectively.  
Therefore, we should expect that these models will not predict the correct gap abundance function, 
$\Phi(\monetwo).$ However, it is still useful to explore the distribution of 
magnitude gaps, given a common richness or distribution of richness.  Such 
a statistic that properly accounts for the influence of the multiplicity function $g(N)$ can serve as a pure test of the ability of the SHAM formalism 
to allocate the brightest galaxies into physically associated, group-sized systems.  

We proceed to perform such a test by randomly selecting a subsample of 
mock groups from our SHAM catalogs with a multiplicity function that matches that 
of the observed, SDSS Mr19 sample.\footnote{Our sample selection proceeds by drawing at random from the mock group sample, assigning a probability to each drawn group of richness $N$ based on $g_{\mathrm{SDSS}}(N)$ and $g_{\mathrm{mock}}(N),$ and then either including or rejecting each drawn group based on the value of an independent draw from a distribution of real numbers uniformly distributed between zero and one.}  Specifically, we compare the observed number density 
of groups as a function of magnitude gap, $\Phi_{\mathrm{SDSS}}(\monetwo)$ to that 
in a subsample of our SHAM mocks restricted to have an identical group multiplicity 
function, $\Phi_{\mathrm{mock}}(\monetwo|g(N)=g_{\mathrm{SDSS}}(N))$.

\begin{figure*}
\centering
\includegraphics[width=8.0cm]{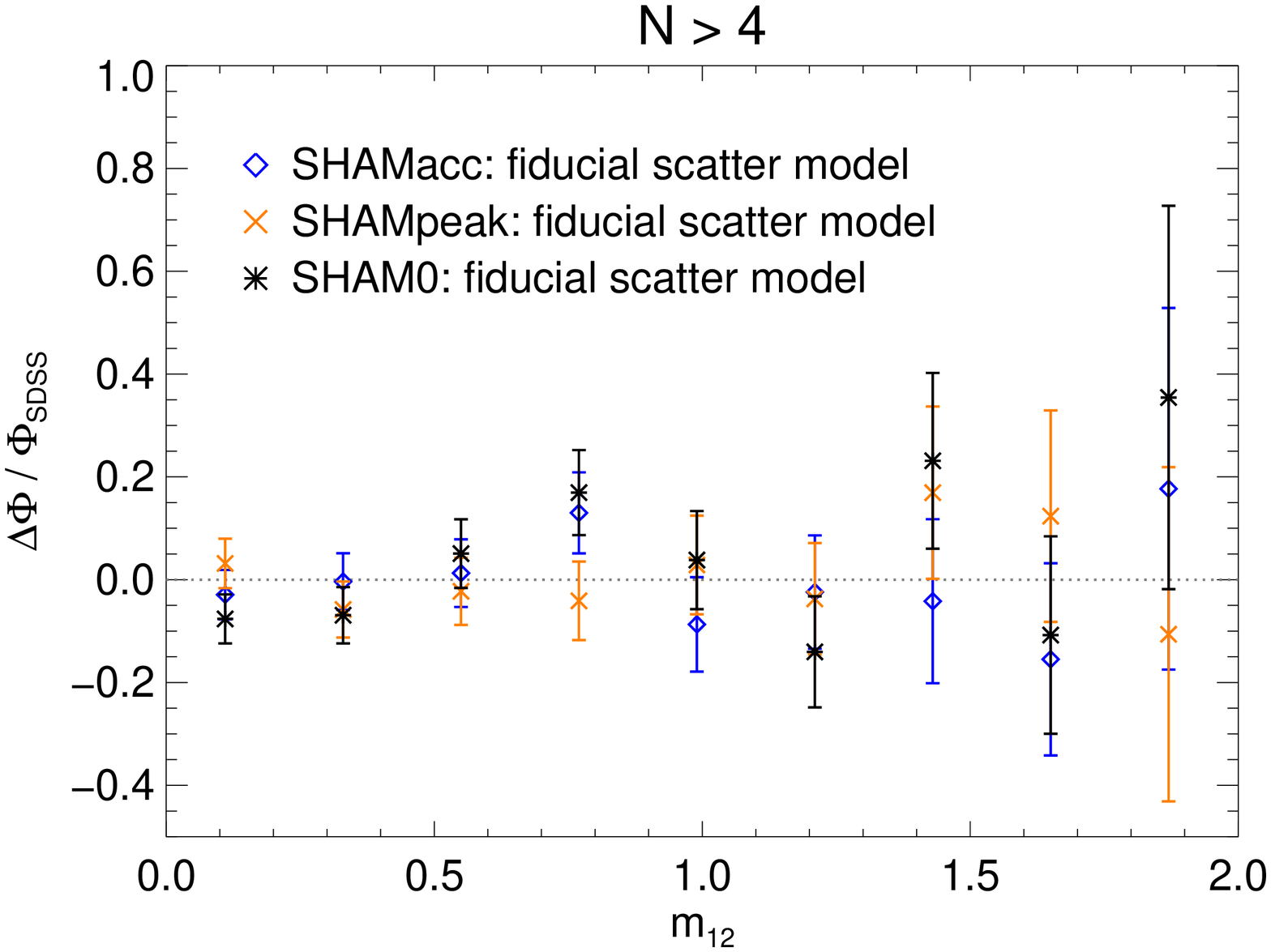}
\includegraphics[width=8.0cm]{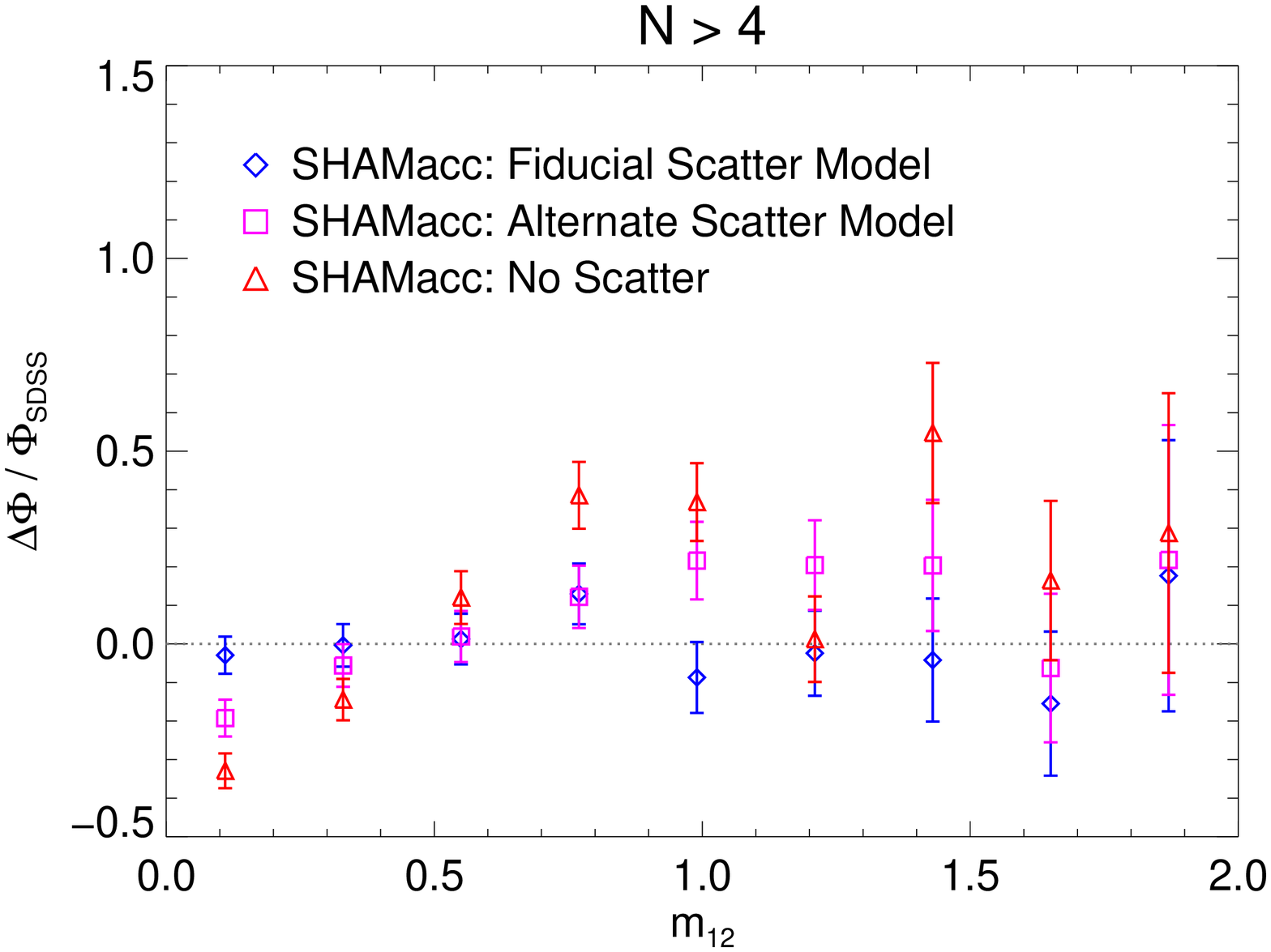}
\includegraphics[width=8.0cm]{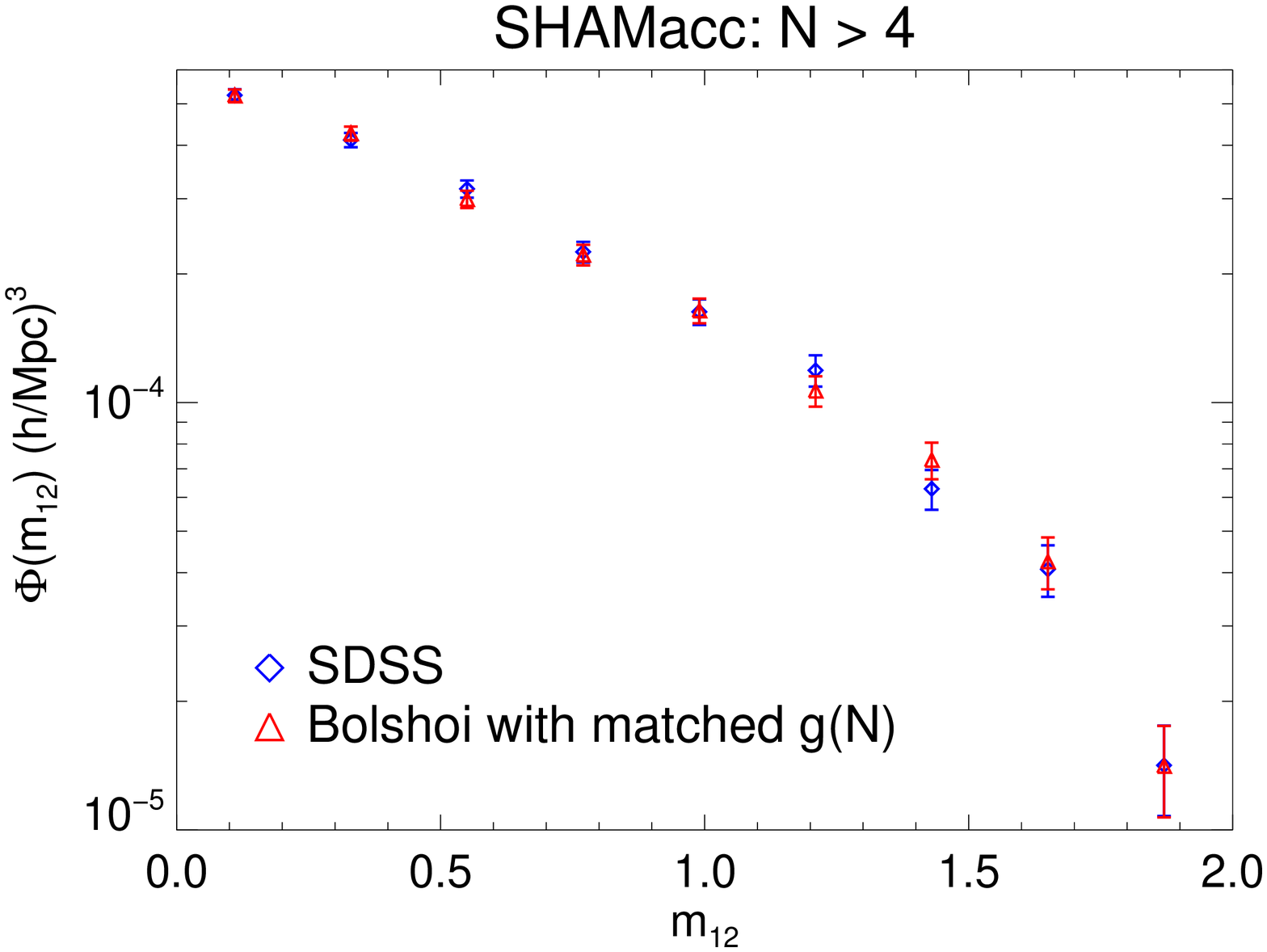}
\caption{
 In the {\em bottom} panel we plot the magnitude gap 
abundance $\Phi(\monetwo)$, exhibited by groups with $N>4$ members. The observed $\Phi(\monetwo)$ is 
illustrated with blue diamonds; we have chosen a random subsample of our fiducial (SHAMacc) mock groups with a multiplicity 
function matching that of the observed groups and plotted the gap abundance exhibited by this 
random subsample with red triangles. 
%The purpose of the multiplicity function matching is to decouple the dependence of the predicted gap abundance from the richness-dependence discussed in \S~\ref{subsubsection:gaprich}. 
The prediction for $\Phi(\monetwo)$ by our fiducial scatter model is less than $1\sigma$ discrepant with the observed distribution, a new success of the SHAM paradigm.
In the {\em top right} panel we show the fractional difference
between the observed gap abundance and that predicted by SHAMacc models for different amounts of 
scatter between $\mr$ and $\vmax$, where $\Delta\Phi\equiv\Phi_{\mathrm{mock}}(\monetwo)-\Phi_{\mathrm{SDSS}}(\monetwo).$
%Our alternate scatter model has a level of scatter that is intermediate between our fiducial model and our mock without scatter. 
%The gap distribution in the alternate scatter model is $2.7\sigma$ discrepant with the observed distribution, while that in the no scatter model greater than $5\sigma$ discrepant, demonstrating the potential to use $\Phi(\monetwo)$ measurements to constrain the scatter in SHAM models.
In the {\em top left} panel we again show the fractional difference between mock and observed $\Phi(\monetwo),$ this time comparing results for mocks where different halo properties were used in the abundance matching. All models shown in the top left panel have the same amount of scatter as that in our fiducial mock catalog. 
%The $\Phi(\monetwo)$ prediction of each three of the models shown in the top left panel is within $2\sigma$ of the observed distribution.
Together with Figure \ref{fig:gn}, these results illustrate that it may be possible to use joint measurements of $g(N)$ and $\Phi(\monetwo)$ to place tight constraints on the details of the SHAM implementation. See text for details.
}
\label{fig:pgap}
\end{figure*}

In the bottom panel of Figure \ref{fig:pgap}, we show the gap abundance $\Phi(\monetwo)$ predicted by 
our fiducial mock catalog and compare it to the observed Mr19 abundance. 
As described above, the subsample of mock groups 
has been chosen to match the observed multiplicity function.  
Thus Fig.~\ref{fig:pgap} illustrates 
the results of a direct test of the abundance matching prediction for the relative 
brightnesses of the two brightest group galaxies, independent of the observed group multiplicity function. 
There is less than a $1\sigma$ difference between the observed $\Phi(\monetwo)$ and our 
fiducial prediction, which constitutes a new success of SHAM-based models for the 
galaxy-dark matter connection.

With the same model of scatter, the matched $g(N)$ 
prediction for $\Phi(\monetwo)$ when abundance matching with either $\vpeak$ or $\vzero$ 
results in less than a $2\sigma$ discrepancy with the data. We show this in the top left panel of Fig.~\ref{fig:pgap}, in which we plot the fractional difference between the predicted and observed magnitude gap abundance for each of the three classes of SHAM models we studied, each with our fiducial model of scatter. 
The sign convention of the fractional difference is given by, $\Delta\Phi(\monetwo) = \Phi_{\mathrm{mock}}(\monetwo)-\Phi_{\mathrm{SDSS}}(\monetwo)$, as in previous plots.  
The magnitude gap is therefore not an effective statistic with which to discriminate between the common choices for the halo property used in the abundance matching algorithm.

For the case of our fiducial mock, we obtain similar results for $\Phi(\monetwo)$ whether or not we employ our $g(N)-$matching procedure. This is because SHAMacc-based mocks correctly predict the group multiplicity function. However, the SHAMpeak and SHAM0 predictions for $\Phi(\monetwo)$ are in stark disagreement with the data
if the comparison is done without first matching the 
predicted $g(N)$ to the observed group multiplicity function, 
demonstrating the importance of multiplicity function-matching when one is interested 
solely in the relative luminosities of the brightest group galaxies.

The success with which SHAM describes the distribution of magnitude gaps 
may seem less interesting because we have already shown that SHAM does not 
accurately predict the luminosities of our group galaxies 
(\S~\ref{subsection:groupfield}, Figs.~\ref{fig:groupfield} \& \ref{fig:groupfieldscatter}). 
However, we reiterate that although these results are related, the gap abundance prediction 
does not follow directly from the group galaxy luminosity function. For example, 
one could have imagined the average luminosities of group galaxies to have 
been {\em correctly} predicted, while the distribution of magnitude gaps was 
incorrect due to a failure of SHAM to correctly arrange bright galaxies within groups. 
In this way, one can see that magnitude gaps test not just the mean conditional 
luminosity function (CLF) of group galaxies, but also correlations between the luminosities 
of group members.

%---------------------------------------------------------------------------------------------------
%-----------------------------------------------------------------------------------------------------

Additionally, magnitude gap abundance observations provide constraints on 
the SHAM models that are {\em complementary} to measurements of group multiplicity. 
We illustrate this in the top right panel of Fig.~\ref{fig:pgap}.  
In that panel, we show the fractional difference between the observed 
abundance of groups as a function of magnitude gap and that predicted by 
SHAM for our different scatter models.  
The red triangles show results from our fiducial scatter model 
($0.2$dex of scatter at the faint end, $0.15$dex at the bright end), 
the magenta squares depict our alternate scatter model (constant scatter of $0.1$dex), 
and the blue crosses show SHAMacc with no scatter.

It is evident that $\Phi(\monetwo)$ is a  
sensitive probe of the underlying scatter between luminosity and $\vmax$.
The alternate scatter model is discrepant with the data at a level of $2.7\sigma$, 
the no scatter model is greater than $5\sigma$ discrepant, and, again, our fiducial model is less than $1\sigma$ discrepant; these results strongly suggesting that $\Phi(\monetwo)$ observations 
can be exploited to constrain the scatter between luminosity and halo circular velocity. 
We reiterate that the gap abundance prediction has been decoupled from the group multiplicity 
prediction for each model, so the complementarity of the constraints on the SHAM model 
provided by $\Phi(\monetwo)$ and $g(N)$ can be realized as these statistics can be used 
concurrently and independently. Magnitude gap or related statistics may thus prove to be useful to constrain scatter in galaxy-halo assignments, 
but we relegate a detailed study of this possibility to future work.

We repeated this entire exercise for mock groups identified in real space 
(as opposed to redshift space), and found that the resulting $\Phi(\monetwo)$ and 
that predicted by our (redshift-space) fiducial mock are different at a level of $\simeq 4.1\sigma$. 
This demonstrates the importance of redshift-space group-finding in making the 
prediction for the gap abundance. 
To our knowledge, we have performed this analysis for the first time; 
all previous studies relying on numerical simulations to predict gap abundances 
have used ``halo-level'' abundances as the prediction, 
in which halo membership is used to define group membership. 
However, real-space predictions systematically under-estimate 
the abundance of low-gap systems found in observations, a fact that we find to hold true 
regardless of the SHAM prescription. This is sensible since interlopers 
occur more often in redshift-space groups, and interlopers can only reduce 
the gap, so it is natural to expect that including interlopers by 
finding the groups in redshift-space should boost the low-gap abundance.  
Since systematic errors due to projection effects play a critical role in group- and cluster-finding, 
{\em any} theoretical study of the magnitude gap abundance must properly account 
for interlopers due to redshift-space projection effects in order to make reliable predictions for magnitude gaps. 
 
 %---------------------------
\subsubsection{Fossil Group Abundance}
\label{subsubsection:ffos}
%---------------------------

\begin{figure}
\centering
\includegraphics[width=8.0cm]{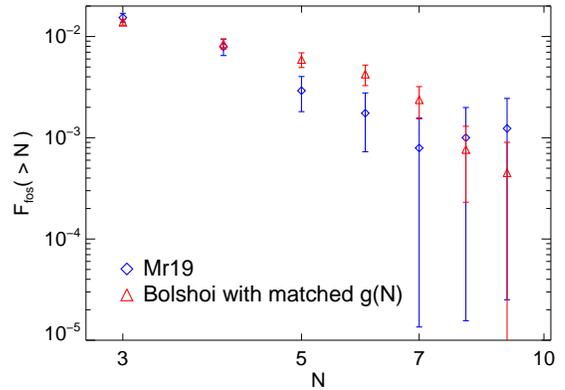}
\caption{
Plot of the {\em fossil fraction} $\ffos(>N)$, defined as the fraction of 
systems in a group sample with magnitude gap $\monetwo\geq 2.$ The richness-threshold defining 
the samples used in the fossil fraction measurements appears on the horizontal axis. 
Blue diamonds (red triangles) show the fossil fraction of the observed (Bolshoi with SHAMacc and a matched $g(N)$) 
group samples for a range of richness cuts. The fossil fraction decreases as the richness threshold defining the group sample increases, another manifestation of the trend of $\monetwo$ with $N$ discussed in \S~\ref{subsubsection:gaprich}.  
Regardless of the richness cut, the SHAMacc prediction for $\ffos(>N)$ is within $2\sigma$ of the observed value, a new success for the abundance matching paradigm.
}
\label{fig:ffos}
\end{figure}

Having studied the full gap distribution in \S~\ref{subsubsection:phigap}, we now focus on the high-gap tail of this distribution.
Plotted as blue diamonds in Figure ~\ref{fig:ffos} is the {\em richness-threshold 
conditioned fossil fraction} $\ffos(>N),$ defined as the fractional abundance of 
SDSS Mr19 galaxy groups with more than $N$ members that have $\monetwo\geq2$: 
\begin{equation}
\ffos(>N)\equiv\frac{\int_{2}^{\infty}\dd\monetwo\Phi(\monetwo|>N)}{\int_{0}^{\infty}\dd\monetwo\Phi(\monetwo|>N)}.
\end{equation}
The fossil fraction is a measure of the size of the large-gap tail of the 
magnitude gap distribution; the decrease of $\ffos(>N)$ with increasing $N$ reflects the relationship 
between richness and gap discussed in \S~\ref{subsubsection:gaprich}: large gap systems are rarer in 
systems of larger richness. Galaxy groups with $\monetwo \geq 2$ whose X-ray brightnesses are 
greater than some threshold value (commonly $L_{X,bol}>10^{42}\mathrm{erg/s}$) 
are often referred to as {\em fossil groups}, and are conventionally thought 
to be galaxy systems that assembled most of their mass at high redshift, 
representing the end products of galaxy group evolution. 
We address the consistency of this interpretation with our findings 
in \S~\ref{section:discussion}. 

Plotted in red triangles in Fig.~\ref{fig:ffos} 
is the SHAMacc prediction for $\ffos(>N)$ from our fiducial mock catalog after 
selecting a random subsample with a $g(N)$ distribution that matches the observed 
Mr19 multiplicity function. For both the mock and observed data points the error bars 
come from bootstrap resampling. In the case of the SHAMacc groups, each bootstrap 
realization corresponds to a new, random selection of a subsample of the mock groups 
with a matched $g(N)$ distribution. Whatever richness cut $N$ one uses to define the group sample, the difference between the observed and predicted $\ffos(>N)$ is less than $2\sigma,$ showing that our fiducial mock accurately predicts the observed abundance of galaxy groups in the large-gap extreme tail of the magnitude gap distribution.

Our approach to both measuring and predicting $\ffos$ differs considerably from much of the existing literature on this subject. In \S~\ref{section:discussion} we discuss how our methods and results compare to previous work. We note here, though, that our methodology for selecting our group sample and measuring magnitude gaps results in the tightest statistical constraints on $\ffos$ in the literature, as well as the best understood systematics, and so the results illustrated with the blue diamonds in Fig.~\ref{fig:ffos} represent the most precise measurement to date of fossil group abundance. For reference, we note that for a group sample with a richness threshold cut of $N>2$ members, we measure the fossil fraction $\ffos(N>2)=0.015\pm0.0016;$ for a richer sample that only includes groups with $N>5$ members, $\ffos(N>5)=0.0018\pm0.001.$

%---------------------------------------------------------------------------------------------------
%-----------------------------------------------------------------------------------------------------

%---------------------------
\section{Tests of Galaxy Group Formation Hypotheses with Data Randomizations}
\label{section:mcs}
%---------------------------

In the previous section, we explored SHAM predictions for a variety of properties of 
galaxy groups.  
As we pointed out in our discussion of Figure~\ref{fig:gaprich} in \S~\ref{subsubsection:gaprich}, 
it is natural to expect the number of group members to be correlated with, for example, 
the luminosity of the brightest group galaxy or the magnitude gap. 
The reason is simple. Consider a toy model in which the luminosities of 
group members are consistent with being random draws from a universal group luminosity function. 
In this case, the typical luminosity of the brightest group galaxy will increase with 
the number of group members because a richer group will draw from the luminosity function more times, 
and hence have more opportunity to include rare, bright galaxies. Likewise, the magnitude gap will decrease with 
the number of group members as the greater number of galaxies will populate the luminosity function more densely (see \S~\ref{subsubsection:gaprich} for a fuller discussion of these trends). 
The fidelity with which such a toy model represents 
the real universe has been explored previously by 
\citet{paranjape_sheth11}, and it is interesting to determine whether 
observed galaxies are consistent with such a simple hypothesis.  

We proceed as follows. For a given set of groups, let 
$\Phi(\monetwo|> N)$ be the abundance of groups of 
a given magnitude gap subject to the condition that 
the group has greater than $N$ members (of course, other conditions 
could be placed on this distribution as well, 
see below for further discussion).  Assume 
some luminosity function of galaxies within such 
groups, $\Phi(L|> N)$.  If we assume that galaxy 
luminosities are consistent with random draws from 
$\Phi(L|> N)$, then under this random-draw 
hypothesis there is a definite prediction for 
the distribution of magnitude gaps, $\phirand(\monetwo|> N),$ given only this luminosity function and the group multiplicity function $g(N).$  Additionally, there is a definite prediction for $\phirand(\li|>N),$ the luminosity function of the $\ith$ brightest galaxies found in groups with more than $N$ members.
Of course, 
real data need not be consistent with this hypothesis, 
so it is interesting to examine deviations from the 
random draw hypothesis.  We do this by defining the 
fractional deviations from the random-draw prediction for gap abundance, 
\beq
\label{eq:adf} 
\Psi(\monetwo|> N) \equiv 
\frac{\Phi(\monetwo|> N)-\phirand(\monetwo|> N)}{\phirand(\monetwo|> N)},
\eeq
and fractional deviations from the random-draw prediction for the luminosity function of the $\ith$ brightest group galaxies, 
\beq
\label{eq:adf2} 
\Psi(\li|> N) \equiv 
\frac{\Phi(\li|> N)-\phirand(\li|> N)}{\phirand(\li|> N)}.
\eeq

Of course, one can make different assumptions about the luminosity 
function from which the galaxies are being drawn in order 
to test different hypotheses. For example, one could draw 
the luminosities from the all-galaxy luminosity function 
$\Phi(L),$ rather than a richness threshold-conditioned LF. 
We intend to explore these and related details in a future follow-up paper. Before presenting our measurements of $\Psi(\li|> N)$ and $\Psi(\monetwo|> N),$ we pause briefly to pinpoint the specific hypothesis that is being tested by comparing the data to this randomization: {\em the physics of group assembly influences the brightness of galaxies in such a way that $\Phi(L|>N)$ is distinct from the global, unconditioned luminosity function, $\phiglo(L),$ but no other imprint is made on galaxy luminosities.}

\begin{figure*}
\centering
\includegraphics[width=8.0cm]{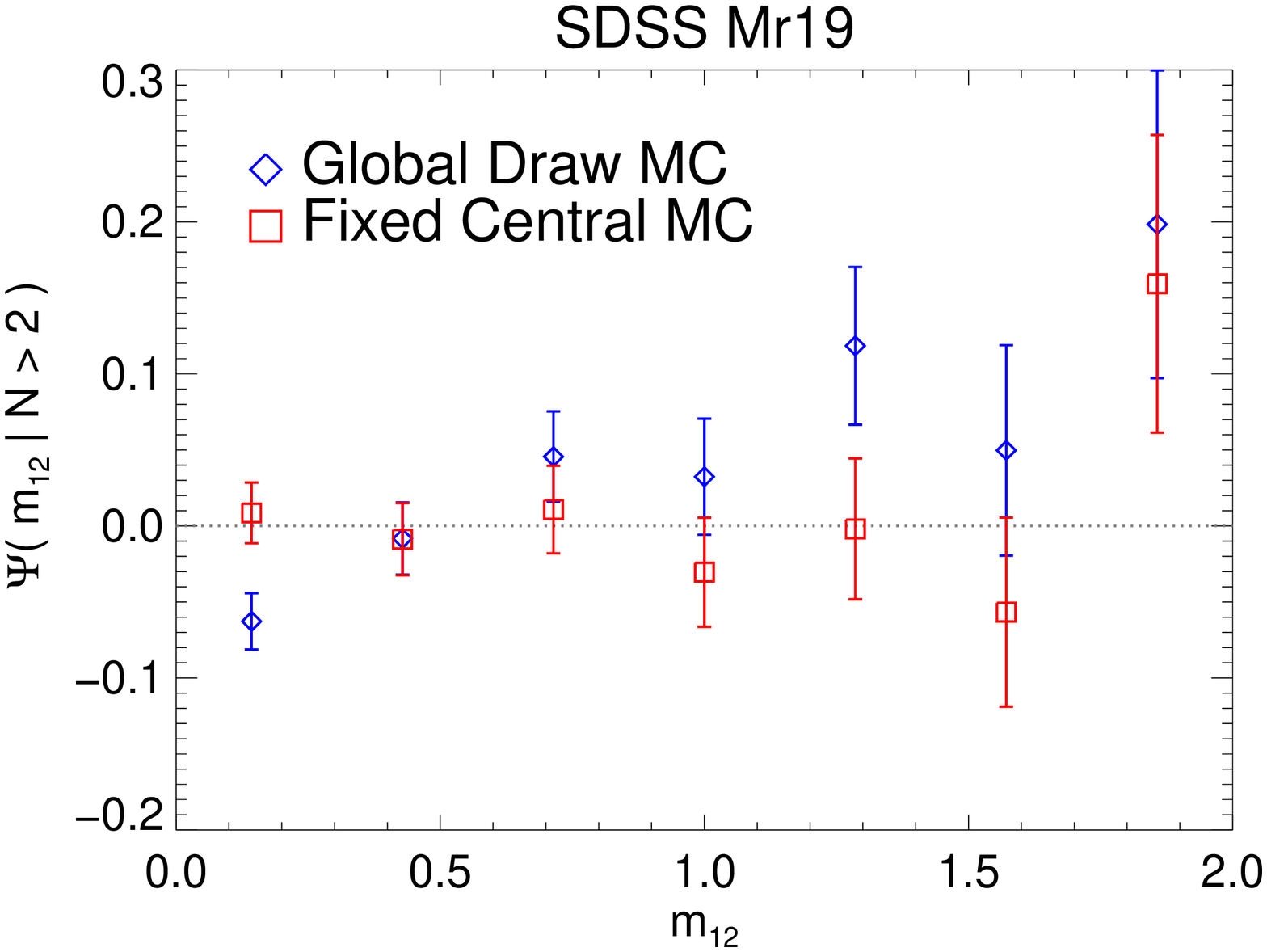}
\includegraphics[width=8.0cm]{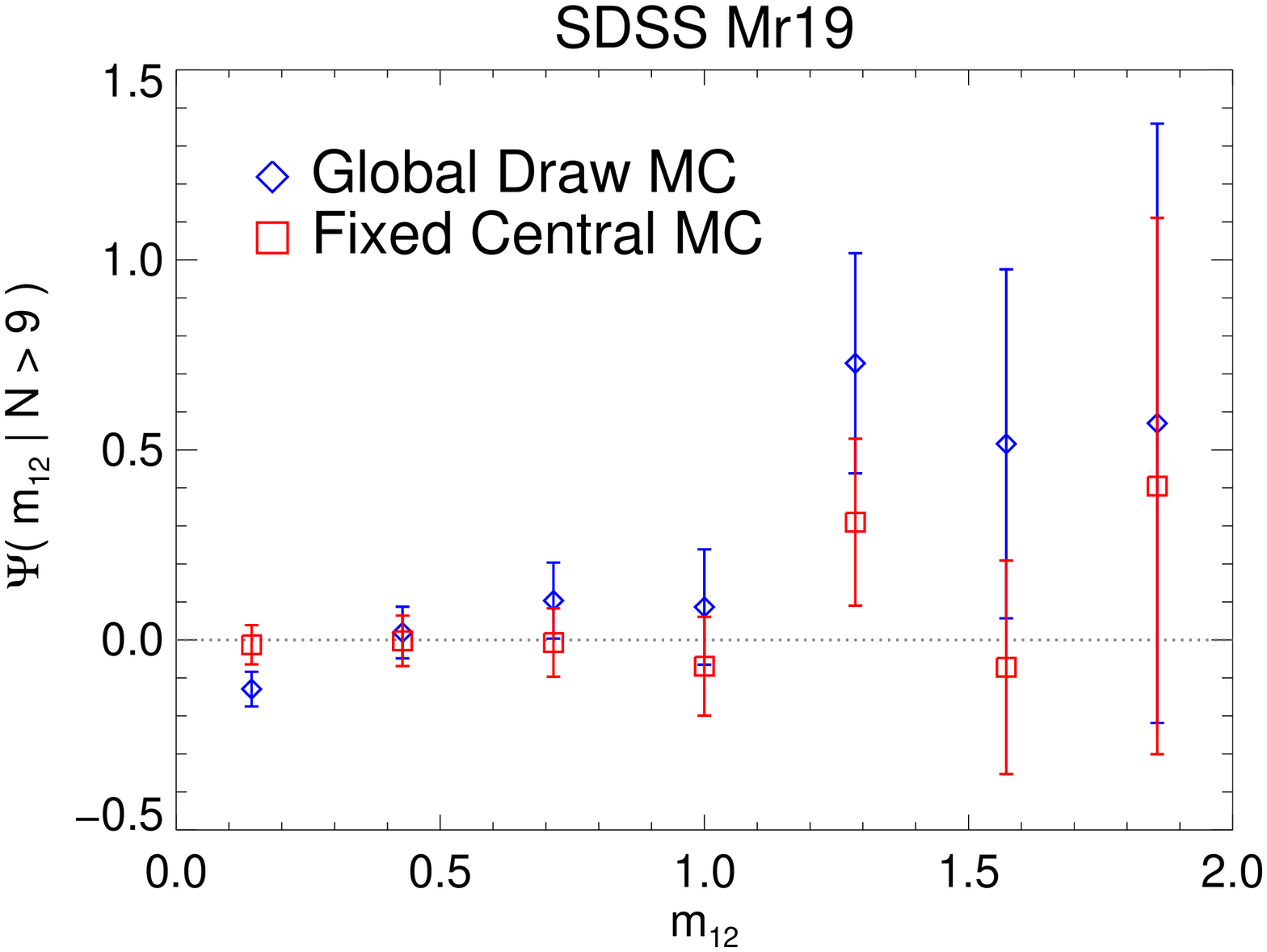}
\includegraphics[width=8.0cm]{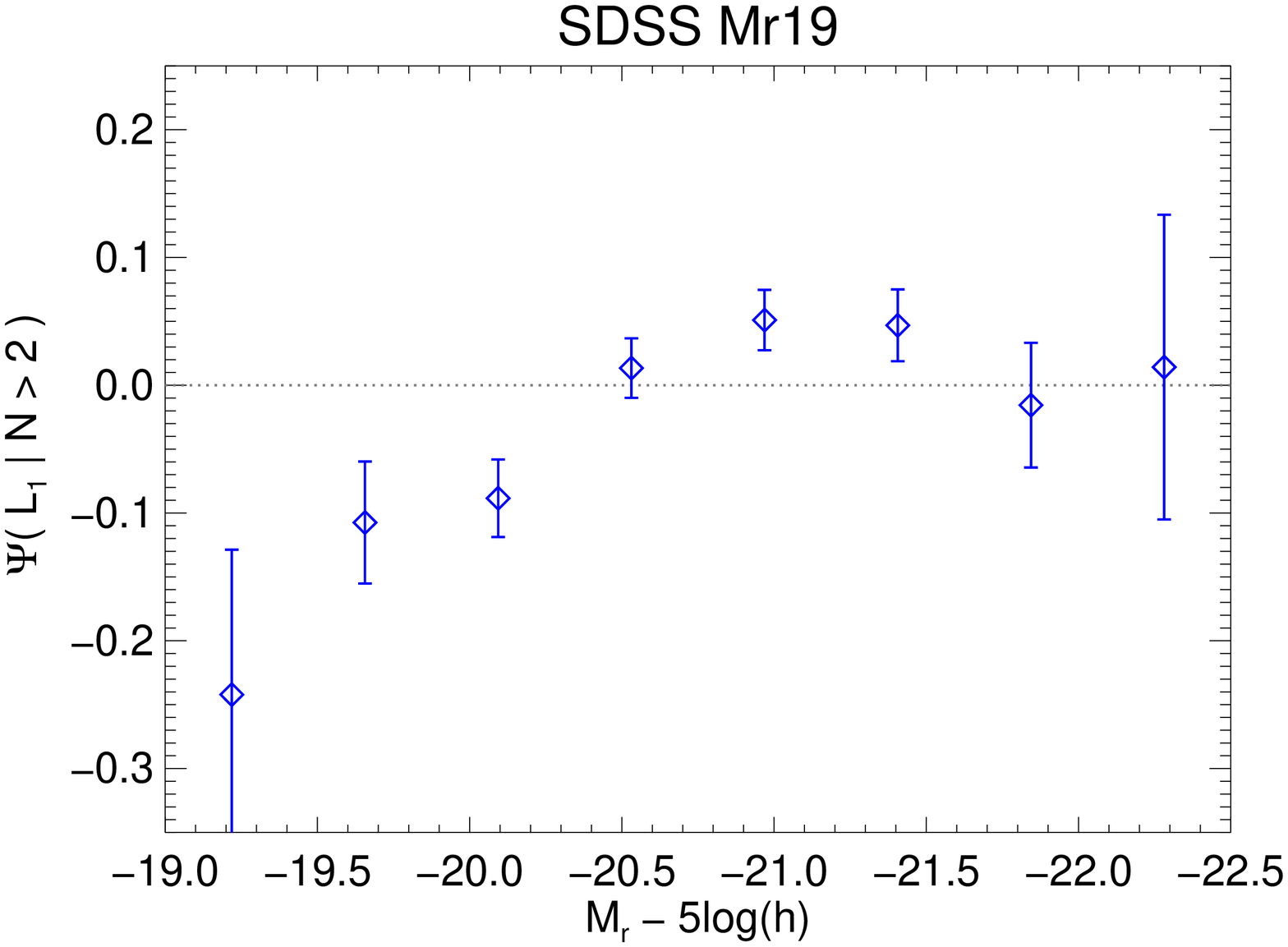}
\includegraphics[width=8.0cm]{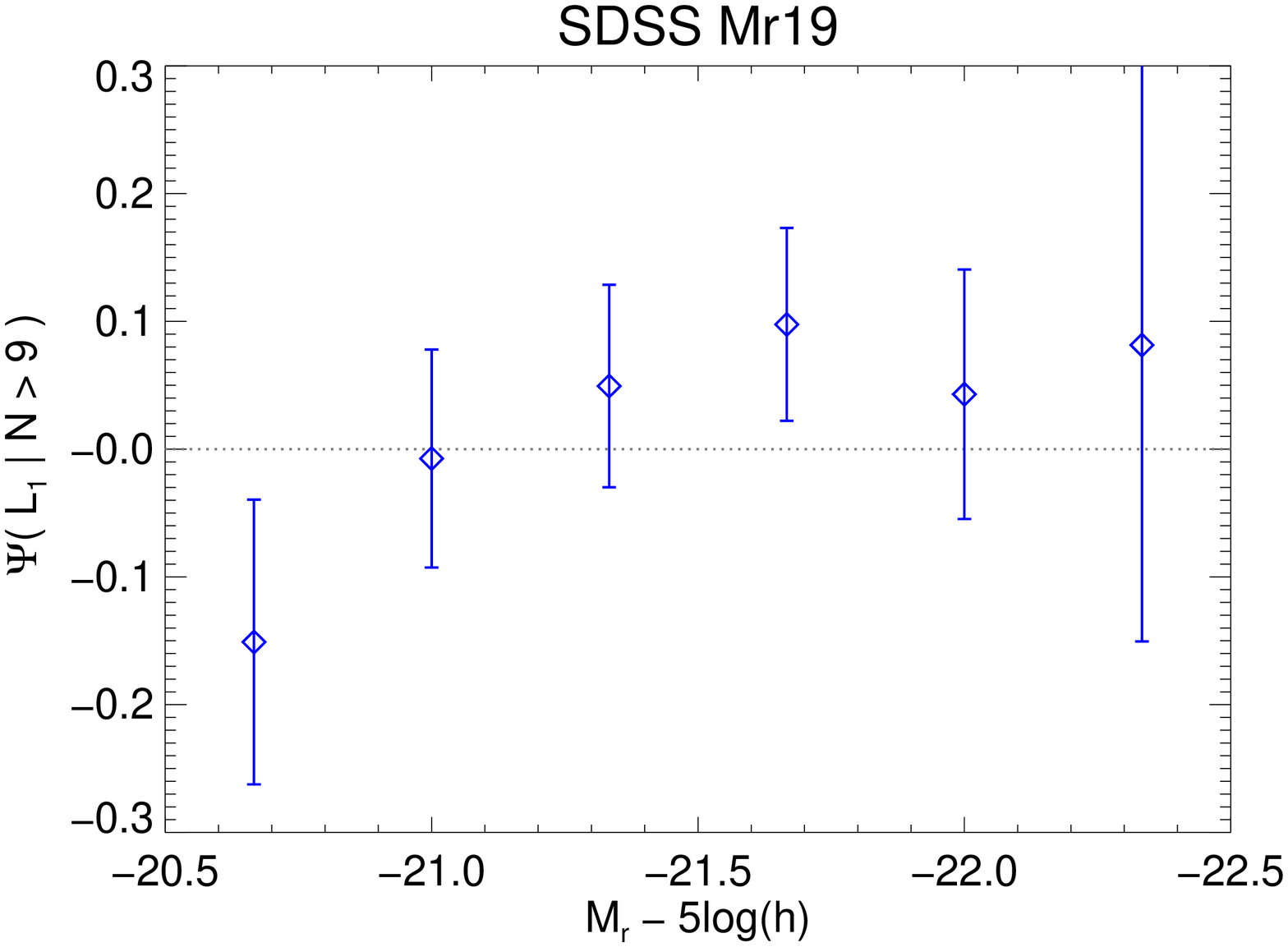}
\caption{
Randomization tests of magnitude-gap statistics and rank-ordered luminosity functions. 
In blue diamonds we plot results pertaining to the data randomization procedure in which we draw brightnesses for
all the group members from the global $\Phi(L|> N).$ With the red squares we plot results for a second data randomization, in which the brightest group galaxy, $L_1,$ is kept fixed and we randomly draw brightnesses for the remaining galaxies in each group from $\Phi_{sat}(L|L_1),$ allowing us to directly test one of the most common assumptions of the Conditional Luminosity Function (CLF) formalism. See text for further
details about the data randomizations. 
In the {\em top panels} we plot  our SDSS measurement of $\Psi(\monetwo|> N)$ (see Eq.~\ref{eq:adf}), the fractional difference between the observed gap abundance and that exhibited by the randomized data, $(\Phi_{\mathrm{SDSS}}-\phirand)/\phirand.$ In the {\em bottom panels} we plot our SDSS measurement of $\Psi(L_1|> N)$ (see Eq.~\ref{eq:adf2}), the fractional difference between the observed luminosity distribution of the brightest group galaxies and that in the randomized data set. 
We show results
for a richness threshold of $N> 2$ in the {\em left panels,} and $N> 9$ in
the {\em right panels.} 
%For the first data randomization (blue diamonds), each of the distributions $\Psi(\monetwo|> N)$ as well as $\Psi(L_1|N>2)$ are distinct from an identically-zero distribution, indicating that the brightness of group galaxies is inconsistent with this simple random-draw hypothesis. This implies that the physics of group assembly leaves an imprint on the brightness of group galaxies that is not encoded solely by group richness. For the second data randomization (red squares), each of the $\Psi(\monetwo|> N)$ are consistent with zero, providing new and direct support for a commonly made assumption in the CLF prescription for constructing mock galaxy samples, namely that it is possible to adequately condition $\phisat(L)$ with knowledge of $\lcen.$
}
\label{fig:psigap}
\end{figure*}

With the blue diamonds in Figure~\ref{fig:psigap}, we plot our measurements of 
$\Psi(\monetwo|>N)$ for SDSS Mr19 groups in the 
top panels, and  $\Psi(L_1>N)$ in the bottom panels; results for groups with $N>2$ members appear in the left panels of Fig.~\ref{fig:psigap}, $N>9$ members in the right panels.  
These results are very similar to the case that 
was tested in \citet{paranjape_sheth11}, except those authors 
used an analytical fit \citep{bernardi_etal10} to $\Phi(L|> N),$ 
and restricted attention to $N>9.$
To make Fig.~\ref{fig:psigap}, we employ different methods that are based on the {\em exact} $\Phi(L|>N)$ exhibited in our sample, which we describe as follows. For each group of richness $\nj>N,$ we construct $1000$ realizations of the group by repeatedly drawing $\nj$ times\footnote{In our data analysis we compute the values of $\monetwo$ and $\li$ in each group by first excluding the group's fiber-collided members (see Appendix B). Accordingly, the random draw predictions should be made with the variable $N_{\mathrm{nfc}},$ the number of non-fiber-collided members of the group.} from the observed $\Phi(L|>N),$ allowing us to calculate $\phirand(\monetwo)$ and $\phirand(\li)$ directly from the randomized data.

The $\Psi(\monetwo|> N)$ plotted with blue diamonds in the top panels of Fig.~\ref{fig:psigap} are 
clearly inconsistent with zero.  The statistical significance 
of this difference is $\simeq 4.8\sigma$ for $N > 2$ groups 
and $\simeq 3.2\sigma$ for $N > 9$ groups.  Alternatively, 
the p-values from two-sided KS tests are $\lesssim 10^{-4}$ 
in both cases. These tests yield the clear conclusion that the allocation of galaxy luminosities 
amongst SDSS groups is not consistent with random draws from a global group-galaxy luminosity function.
This is a direct contradiction of the conclusions drawn 
in \citet{paranjape_sheth11} based on the same statistical test. The primary source of the discrepancy between 
these two conclusions lies in the treatment of fiber collisions, 
which we provide a detailed account of in Appendix B. 

As for the results plotted in the bottom panels, we find that $\Psi(L_1|N>2)$ is distinct from zero at a level of $4.0\sigma$, but that $\Psi(L_1|N>9)$ is consistent (within $1\sigma$) with zero. Thus, our results support the conclusion of \citet{paranjape_sheth11} that, in groups with $N>9$ members, the luminosity function of the brightest group galaxies is well described by $\phirand(L_1|N>9);$ as pointed out in \citet{more12}, the sample size of groups with $N>9$ members is quite limited, and so only quite strong differences would be evident in this sample. Evidently, however, in groups with only a few members some additional variable besides the richness of groups is clearly necessary to encode the influence that the physics of group assembly has on the luminosities of the brightest group members. We further discuss the interpretation and consequences of this result in \S~\ref{section:discussion}.

We have also measured the $\Psi(\monetwo|> N)$ that results from 
an alternative data randomization procedure, which we describe as follows. 
First, we divide our SDSS Mr19 sample of galaxies into ``centrals'' and ``satellites''; 
the set of centrals is defined to be those galaxies that are the brightest among 
the galaxies in the group of which they are a member; the set of satellites is 
the complement to the set of centrals. For each group of richness $N$ in the 
sample being randomized, we construct $1000$ realizations of the group. For each 
realization, we fix $L^{\mathrm{ran}}_1,$ the luminosity of the ``first'' randomized group member, 
to be equal to $L_1,$ the luminosity of the central galaxy in the group that 
is being randomized. The luminosities of the remaining $N-1$ members are drawn 
at random from $\Phi_{sat}(L|L_1),$ the luminosity function of all satellite 
galaxies that are found in groups whose central galaxy has a luminosity within $0.2$dex of $L_1.$

This randomization scheme may appear unfamiliar at first glance, but in fact 
it is well-motivated by and quite similar to the most commonly used method by which mock catalogs of 
galaxies are constructed from N-body simulations in the standard Conditional Luminosity 
function (CLF) formalism \cite[e.g.,][]{yang_etal03,yang_etal04,more12}.  CLF-based mocks distinguish between central and satellite galaxies, although the distinction is not defined by a rank-ordering of galaxies by brightness. Instead, the brightness of the central galaxy, $\lcen,$ is assumed to be drawn at random from 
a log-normal luminosity function whose mean and spread are governed by the mass of the 
dark matter halo, and the luminosities of the remaining galaxies associated with the halo 
are chosen from a modified Schechter function whose form is governed by $\lcen.$ 

Our approach to the construction of our randomized groups is very similar except that (1) we know the richnesses 
of the groups rather than their masses, and (2) we suppose that the brightest galaxy in the group has the same luminosity as the galaxy embedded in the group's most massive halo, that is, $L_1\approx\lcen.$ By keeping fixed the luminosity of the central galaxy 
in each randomized group, our randomization preserves the relationship between $\lcen$ and $N$ 
exhibited by the observed groups. Rather than using an analytical fit to $\phisat,$ we 
use the data itself to determine this distribution; our choice to condition $\phisat$ with the luminosity of the central galaxy  
mirrors the convention commonly utilized in standard implementations of the CLF formalism. 

A non-negligible fraction of the randomly-drawn satellites prove to be brighter than 
the central galaxy in the randomized group (as should be expected from the results in \citet{skibba_etal07}). However, in calculating the magnitude 
gap of each randomized group we proceed exactly as we do with the data and 
define $\monetwo$ to be $M_{r,1}-M_{r,2},$ the magnitude difference between the 
brightest two members. The resulting $\Psi(\monetwo|> N)$
is plotted with red squares in the top panels of Fig.~\ref{fig:psigap}.\footnote{We do not include plots from this data randomization in the bottom panels since keeping $\lcen$ fixed results in the trivial prediction that $\Psi(L_1|>N)=0.$} Regardless of the richness cut, this second data randomization faithfully reproduces  $\Phi(\monetwo|>N)$ (by construction, $\Phi(L_1|>N)$ is exactly reproduced in this randomization). This agreement provides new supporting evidence of a common, underlying assumption of the CLF prescription for constructing mock galaxy samples, namely that $\phisat$ need only be conditioned on $\lcen$ in order to accurately account for the luminosity function of satellite galaxies. 

%---------------------------------------------------------------------------------------------------
%-----------------------------------------------------------------------------------------------------

%--------------------------
\section{Discussion}
\label{section:discussion}
%---------------------------

%--------------------------
\subsection{Conditional Luminosity Function Results}
\label{subsection:clfdiscuss}
%---------------------------

 %--- Discussion of group and field LF results

We have used a volume-limited catalog of galaxy groups observed in
SDSS DR7 to provide a number of new tests of the abundance matching
prescription for connecting galaxies to dark matter halos. To perform these tests, we have developed a novel implementation of SHAM that allows for the
construction of mock galaxy catalogs with a luminosity function
$\Phi(L)$ that exactly matches the $\Phi(L)$ exhibited by an 
observed galaxy sample, even when scatter between halo circular velocity 
and galaxy luminosity is included. We have
exploited this implementation to test the ability of SHAM to predict
the galaxy luminosity function conditioned on whether the galaxies in
the sample are members of groups. 
We find that field (group) galaxies in SHAM-based catalogs are systematically 
too dim (bright), with differences at the level of $\gtrsim10\%$ at virtually all luminosities, and that this behavior holds true in all of the 
SHAM models we studied. This indicates that none of the widely-used  
SHAM-based models allocate galaxies to field and group environments 
to better than $\sim10\%$ accuracy. These findings are particularly interesting in the context of \citet{neistein_etal11}, whose results suggest that subhalo environment may need to be included in the abundance matching prescription in order to accurately model galaxy clustering. 

In a study closely related to ours, \citet{reddick_etal12}
construct SHAM-based mock catalogs of galaxy groups (using a different implementation of the abundance matching prescription) and studied a wide range of SHAM-based models of the galaxy-halo connection, so it 
is useful to compare our results with theirs.  For detailed CLF comparisons we rely on private communications with Rachel Reddick, because our group-finding algorithm differs from theirs in that ours does not have group masses intrinsically built into it, and \citet{reddick_etal12} quote  CLF results that are conditioned on group mass. Both the field and group galaxy luminosity function of their best-fit model is discrepant at the $\sim20\%$ level with the data, a level of disagreement that is comparable to ours. However, the character of the failure apparently depends on both the SHAM implementation as well as the group-finding algorithm. This emphasizes a need for systematic and detailed examinations of the influences of mock catalog construction, group finding, and other methodological issues in order to understand the potential systematic differences induced by these choices. We intend to explore these issues in future work.

%One difference between our mock catalogs and those in \citet{reddick_etal12} is that their SHAM implementation begins by abundance matching to the exact luminosity function of the Mr19 sample, $\Phi_{\mathrm{SDSS}}(L),$ but their iterative procedure for adding scatter results in a final luminosity function that is $\sim1\sigma$ discrepant with $\Phi_{\mathrm{SDSS}}(L).$ Thus it is possible that this discrepancy serves to mask the true difference between the observed $\Phi_{group}(L)$ and that predicted by their model. There are few additional possibilities that may account for the differences between our results and those of \citet{reddick_etal12} because we analyze the same N-body simulation, using the same halo-finder, and compare to the same galaxy sample. The primary differences between our methodologies lie in the details of our SHAM models and in our group-finding algorithms. If either of these two differences accounts for the different luminosity trends, it would have important consequences for the construction of mock galaxy catalogs that faithfully represent the properties of observed galaxies. This emphasizes a need for systematic and detailed examinations of the influences of mock catalog construction, group finding, and other methodological issues in order to understand the potential systematic differences induced by these choices. 

 %--------------------------
\subsection{Group Multiplicity Function Results}
\label{subsection:gndiscuss}
%---------------------------
%--- Discussion of g(N) results

In \S~\ref{subsection:gn} we demonstrated that our fiducial mock galaxy
catalog, constructed by abundance matching using $\vacc$ as the luminosity proxy for subhalos, 
accurately reproduces the observed group multiplicity function $g(N),$ that is,
the abundance of groups as a function of group richness. Furthermore, in \S~\ref{subsection:gn} we also showed that mock catalogs using $\vpeak$ (SHAMpeak models) or $\vzero$ (SHAM0 models) as the abundance matching parameters incorrectly predict group multiplicity
measurements and straddle the $g(N)$ predicted by $\vacc-$based
catalogs (Figure ~\ref{fig:gn}). We have checked that this qualitative
behavior holds true for models with very different (or without)
scatter between luminosity and $\vmax,$ as well as for volume-limited
samples with different brightness thresholds, 
indicating that this a generic conclusion. 

The multiplicity function prediction constitutes a new success of SHAM that is distinct from previous tests
that rely on measurements of galaxy clustering.\footnote{See also \citet{skibba_sheth09} for an alternative approach to probing the galaxy-halo connection beyond galaxy clustering.} First, we note that because group identification occurs in redshift-space, $g(N)-$based tests are  sensitive to the Bolshoi prediction for the velocity field, whereas measurements of  (projected) galaxy clustering, $w_{\mathrm{p}}(r_{\mathrm{p}}),$ purely probe the prediction for the spatial distribution of halos. This  difference alone distinguishes our $g(N)-$tests from tests of SHAM based on clustering. However, with Fig.~\ref{fig:xin} we demonstrated an additional difference between the two tests: galaxy clustering measurements on small scales are insensitive to differences in $g(N)$ in the $N\gtrsim20$ regime. 

This second difference has important implications for conventional tests of theories predicting how galaxies populate dark matter halos. While we find that no SHAMpeak model provides an adequate description of the group multiplicity function in the $N\gtrsim20$ regime, \citet{reddick_etal12} showed that all SHAMacc models under-predict $w_{\mathrm{p}}(r_{\mathrm{p}})$ on small scales\footnote{We have independently verified that this behavior occurs in our mock catalogs, and so we confirm the conclusion drawn in \citet{reddick_etal12} that SHAMpeak models provide a more accurate prediction for small-scale galaxy clustering than SHAMacc models.}  ($r_{\mathrm{p}}\lesssim1\hmpc$). However, as we showed in Fig.~\ref{fig:xin}, the small scale clustering measurements are insensitive to $g(N)$ measurements in the $N\gtrsim20$ regime. 

Taking these results together, we tentatively conclude that {\em no SHAM model studied in either of these works can simultaneously predict small-scale galaxy clustering and the group multiplicity function correctly}. While an exhaustive exploration of the parameter space used in the SHAM implementation would be required to conclusively establish this result, there is little room to ameliorate this mutual inconsistency by adjusting the amount of scatter in the SHAM implementation: the scatter between $\mr$ and $\vmax$ must be similar to that in our fiducial model in order to correctly predict $\Phi(\monetwo),$ and the SHAM prediction for $g(N)$ is largely insensitive to the amount of scatter. Moreover, by repeating our analysis on catalogs constructed via the \citet{reddick_etal12} SHAM algorithm (Rachel Reddick, private communication), we have shown that our multiplicity function results persist regardless of the details of the abundance matching implementation, including their modeling of tidal disruption. This conclusion is thus likely to be robust to any conventional refinement of SHAM.\footnote{Although see \S~\ref{subsection:orphans} for an important caveat related to orphan galaxies.} 

It is interesting to consider these results in the context of Halo Occupation Distribution (HOD) models, which describe the probability that a halo of mass $\mh$ is populated with $N$ galaxies of some type, $P(N|\mh).$ Since small-scale galaxy clustering is dominated by halos hosting only a few galaxies, and since such halos are typically smaller in mass relative to halos that host a large number of galaxies, the $w_{\mathrm{p}}(r_{\mathrm{p}})-$based results in \citet{reddick_etal12} demonstrate that SHAMpeak models provide a more accurate description of the HOD at the low-mass end. 

On the other hand, since dark matter halos hosting $N\gtrsim20$ galaxies are typically very massive, our $g(N)-$based results demonstrate that SHAMacc models provide a more accurate description of the HOD at the high-mass end. Taken together, these results illustrate that {\em measurements of the group multiplicity function and small-scale galaxy clustering probe different regions of HOD parameter space}, in keeping with the findings in \citet{berlind_weinberg02}. It would therefore be very interesting to constrain HOD models using {\em joint} measurements of $w_{\mathrm{p}}(r_{\mathrm{p}})$ and $g(N),$ although modeling the covariance between these observables would be highly non-trivial. Additionally, such an undertaking would require including the effect that cosmology has on $g(N),$ particularly in the high mass (large richness) regime. We leave the investigation of this possibility as a task for future work.

The possibility remains that both $g(N)$ and $w_{\mathrm{p}}(r_{\mathrm{p}})$ could be correctly predicted together by some previously unexplored hybrid between the two models, perhaps in which subhalos are assigned luminosities according to either $\vacc$ or $\vpeak,$ depending on the mass of their host halo. While such a model would be quite contrived from a theoretical standpoint, it may nonetheless prove useful in the construction of mock galaxy catalogs. We leave the development and exploration of such a model as a task for future work.

Our multiplicity function results are also intriguing in the context of a recent study by
\citet{watson_etal11}, who showed that incorporating stellar mass loss
into SHAM-based models of galaxy formation improves the predictions
for galaxy clustering. Since a subhalo's $\vmax$ value at the time of
accretion represents an intermediary stage of dark matter mass loss
between $\vpeak$ and $\vzero,$ our results point towards the
possibility that $g(N)$ measurements may have the potential to
constrain evolution of stellar mass (both new star formation and 
stellar  mass loss) in satellite galaxies within group and cluster halos.

 %--------------------------
\subsection{Magnitude Gap Results}
\label{subsection:pgapdiscuss}
%---------------------------
 %--- Discussion of magnitude gap results
One of the most common statistics used to quantify magnitude gap abundance is
the {\em fossil fraction}, $\ffos,$ defined as the fraction of galaxy groups in
a given sample with a magnitude gap $\monetwo \geq 2$.  This is the
statistic we present in Figure
\ref{fig:ffos}. Our methodology for measuring $\ffos$ differs considerably from that which has been adopted in much of the literature on fossil groups. Most significantly, our group sample has not been limited by an X-ray threshold. Thus it may not be surprising that our measurement of the fossil fraction is quite different from the commonly quoted $8-20\%$ value in \citet{jones_etal03}. 

In their study of the Millennium Gas Simulation, \citet{dariush_etal07} studied the influence of the X-ray cut on the magnitude gap distribution, finding that fossil groups that meet the conventional X-ray threshold requirement do not represent a distinct class of objects. They also found that the variance about the mean $\monetwo$ value of a group sample steeply increases as the X-ray threshold defining the sample is relaxed. 

Taken at face value, these results imply that the primary advantage in employing an X-ray threshold is a simple reduction in the intrinsic noise in the measurement. On the other hand, abandoning the X-ray threshold radically increases the sample size of the groups, which, as evidenced by Fig.~\ref{fig:ffos}, results in far more precise measurements of $\ffos$ than that which can be obtained from existing X-ray group samples. While there has been considerable recent progress towards understanding the influence of an X-ray brightness cut on galaxy group properties, \cite[e.g.,][]{george_etal11}, X-ray brightness selection effects remain to be a significant source of systematic uncertainty in any measurement of the cosmic abundance of magnitude gaps based on X-ray samples. This is the primary reason we advocate using an optical galaxy sample in any detailed study of magnitude gaps based on currently available data.

Among the existing
results in the literature, the approach taken in \citet{yang_etal08} and \citet{vandenbosch_etal07} is the most
similar to ours: the authors employ their group-finding algorithm on a
volume-limited, optical sample of galaxies and simply define
$\monetwo$ to be the difference in r-band magnitude between the
brightest two group members. Thus their definition of a fossil is very
similar to ours, and they quote fossil fractions for several different
ranges of group mass, ranging from $0.5\%$ for groups of mass
$\sim10^{14.5}\msun$ to $18-60\%$ for groups of mass
$\sim10^{13}\msun.$ Unfortunately, the mass-binning of these values
makes a direct, quantitative comparison impossible because, unlike the
group-finding algorithm in \citet{yang_etal08}, our algorithm does not
enforce the same assumptions about dark matter halo properties, 
and so groups found with our algorithm are not
inextricably connected with a unique prediction for group
mass. Nonetheless, the fossil fractions quoted in \citet{yang_etal08}
do appear to be significantly higher than those we report in
Fig.~\ref{fig:ffos}, which may be another sign of the sensitivity of
group properties to the algorithms with which they are identified. However, we reiterate a point first made in \citet{paranjape_sheth11} that we have elaborated upon in \S~\ref{subsubsection:gaprich}: the natural statistical correlation between group richness, $N,$ and $\monetwo$ implies that $N$ is the appropriate variable to use to condition the group sample in studies of magnitude gaps, not group mass.

Our study of the magnitude gap statistic is closely related to two
recent investigations of the properties of brightest group galaxies
predicted by the CLF \citep{more12,skibba_etal11b}. Figure 3 of
\citet{more12} demonstrates quite clearly that $\Phi(\monetwo)$ is a sensitive probe of the properties of the satellite galaxy luminosity function. Meanwhile, \citet{skibba_etal11b} showed that the properties of $\Phi_{\mathrm{sat}}(L)$ strongly influence $f_{\mathrm{BHNC}},$ the fraction of brightest-halo galaxies that are not central galaxies, a statistic very closely related to $\monetwo.$ These results suggest that the magnitude gap statistic can be exploited to provide constraints on the treatment of $\Phi_{\mathrm{sat}}(L)$ in the CLF. We intend to explore this possibility in future work.

Recently, \citet{proctor_etal11} claimed that the low richness of the
ten fossil groups they studied indicates a problem for the standard
scenario of fossil group formation. In particular, \citet{proctor_etal11} 
argued that the fact that their fossil groups are under-rich at all 
observed luminosities is difficult to understand within the standard theoretical framework for the formation and evolution of fossil systems.  Our results are a direct
refutation of this claim. First, one should {\em expect} that groups with a large magnitude gap should typically be low-richness systems, an expectation that is highlighted in Fig.~\ref{fig:gaprich}. 
Second, the magnitude gap abundance and fossil
fraction exhibited by our fiducial mock catalog {\em constitutes} the
simplest prediction for these statistics that is consistent with the standard theoretical framework for predicting fossil abundance, namely $\lcdm.$ We have shown that 
the predictions of this standard framework are in good agreement with the data. 
We thus find no evidence that the abundance or properties of fossil systems 
presents a problem for the $\lcdm$ picture of structure formation.

%--------------------------
\subsection{Orphan Galaxies}
\label{subsection:orphans}
%---------------------------

%
Orphan galaxies are defined as galaxies occupying dark matter halos with a mass that has fallen below the resolution limit of the simulation; orphans are typically associated with subhalos that have experienced strong tidal forces through their orbital history inside the virial radius of their host halo. The influence of the orphan population has not been 
accounted for in the construction of our mock catalogs, but may be relevant to some of our conclusions concerning the SHAM model that is preferred by observations. In this section we discuss how orphans may influence our results and conclusions.

Recently, \citet{guo_etal11} applied a semi-analytic model of galaxy formation to the Millennium Simulation to demonstrate that the inclusion of orphan galaxies is required to accurately predict the number density profile of galaxies in clusters. Another related result appears in \citet{watson_etal11}, who showed that on small scales the two-point correlation function is affected at the $\sim20\%$ level when the $\vmax$ completeness limit of a simulation changes from $80$ km/s to $20$ km/s (for reference, the Bolshoi simulation is complete to $\sim53$ km/s. 
\citet{reddick_etal12} found that the satellite fraction predicted by SHAMacc models is too small to correctly predict small-scale clustering, and used these findings to rule out SHAMacc models in favor of SHAMpeak models. However, since \citet{reddick_etal12} did not model the orphan population, and since including orphans will increase the number of subhalos in the mock catalog and thereby boost the satellite fraction, it is possible that the inclusion of an orphan population would bring SHAMacc models into better agreement with galaxy clustering measurements.\footnote{Although see Appendix B of \citet{reddick_etal12} for their tests of resolution effects.} These results motivate the treatment of orphans in precision studies of the galaxy-dark matter connection, and suggests that small-scale clustering may also be sensitive to the orphan population. 

One approach to this problem appears in \citet{moster_etal10}, who model the effect of orphan galaxies by following the evolution of the most-bound particle in the subhalo after the subhalo falls below the resolution limit of the simulation. Alternatively, an orbital evolution code \citep[e.g.,][]{zentner_etal05} might be employed for the same purpose. However, comparative tests of the consequences of different  modeling choices would need to be conducted before conclusive statements could be made concerning the true importance of the impact of the orphan population. Such an investigation is beyond the scope of the present work, but we intend to explore this possibility in a future paper.

The boost to the satellite fraction provided by modeling orphan galaxies may also influence the group multiplicity function since this would generally lead to an increase in group richness. However, since SHAMpeak models already over-predict $g(N\gtrsim20),$ this effect is unlikely to change our qualitative conclusion that our $\vpeak-$based mocks generically fail to correctly predict the abundance of rich groups.  

Including the orphan population in our mocks will tend to boost $\Phi_{\mathrm{group}}(L),$ particularly at the faint end, resulting in a concomitant decrease in faint field galaxies. As can be seen by inspection of Fig.~\ref{fig:groupfield}, this is the correct sense of the change to $\Phi_{\mathrm{group}}(L)$ that would be necessary to bring the SHAMacc prediction into agreement with the data, but it appears unlikely that agreement between the SHAMpeak predictions and the data could be brought about by such a change. 

Finally, we consider it unlikely that luminosity gap statistics will be strongly affected by the treatment of orphans because this statistic is dominated by the brightest galaxies in the group environment, whereas including orphans will primarily influence the faintest galaxy population. 

%We conclude this section by noting that orphans have not been modeled in {\em any} previously studied SHAM-based mock catalog.\footnote{Although see \citet{moster_etal10} for an alternative method of constructing mock galaxy populations that includes orphan galaxies.} Thus, while modeling this population may have an impact on some of our results, we have nonetheless performed a thorough exploration of all previously studied classes of mock catalogs based on abundance matching. We leave the detailed modeling of orphan galaxies as a task for future work.

%--------------------------
\subsection{Data Randomization Results}
\label{subsection:mcdiscuss}
%---------------------------

Data randomization techniques similar to the ones we use in
\S~\ref{section:mcs} have been used previously in the literature. For
example, \citet{jones_etal03}, \citet{dariush_etal07}, and
\citet{tavasoli_etal11}, all addressed the connection between richness
and magnitude gap with Monte Carlo (MC) realizations of a group
sample.\footnote{See also \citet{vale_ostriker08} and references therein for early attempts to use the magnitude gap and other statistics to characterize the ``specialness" of the central galaxy luminosity.} These authors constructed a population of $10^{4}-10^{6}$
Monte Carlo groups by drawing a fixed number of times from a global
Schechter luminosity function to populate each MC group with a set of
galaxies. In finding that the fraction of their MC groups with
$\monetwo \geq 2$ was lower than that of the groups in their sample,
they each concluded that fossil groups do not 
have a ``statistical origin.''  The authors interpreted 
these exercises as evidence that fossil groups do not arise 
as extreme realizations of a Poisson process based on the global 
galaxy luminosity function, but through a 
dynamical process that preferentially eliminates satellite 
galaxies with large luminosities, namely mergers driven 
by dynamical friction.

More recently, \citet{paranjape_sheth11} applied a statistical technique first introduced by \citet{lin_etal10} to the galaxy sample we study in this work to
construct a Monte Carlo realization of a group sample with the same
multiplicity function as the observed group sample, finding that
$\Phi(\monetwo),$ the abundance of groups as a function of magnitude
gap, is well-described by their Monte Carlo population. As discussed
in \S~\ref{section:mcs} and in Appendix B, we repeat the
\citet{paranjape_sheth11} analysis with an improved treatment of fiber
collisions and reach the opposite conclusion. 
We find that the distribution of magnitude gaps is inconsistent with galaxy luminosities being drawn randomly from a group luminosity function, $\Phi_{\mathrm{group}}(L|>N).$

Tests based on the Monte Carlo realizations described above, including our own, can only determine
whether or not knowledge of the richness of groups together with a
universal galaxy luminosity function provides sufficient information
to predict the magnitude gap distribution. In \S~\ref{section:mcs} we
established that such information is insufficient. However, from this
we can only conclude that knowledge of some other group property
besides richness is required to predict the observed $\Phi(\monetwo)$. 
This insufficiency does not, taken by itself, reveal the physical origin of
systems with a large gap.

However, our mock galaxy catalogs provide more compelling evidence in favor of the dynamical picture of fossil group origins.
 The cosmological simulation on which our mock
catalog is based traces the evolution of a $\lcdm$ universe from the
initial seeds of structure formation through to the present day,
including the dynamical processes conventionally thought to
determine the magnitude gap. The successful prediction for
$\Phi(\monetwo)$ of our fiducial mock catalog thus provides strong
supporting evidence that the magnitude gap exhibited by galaxy groups
is influenced by dynamical processes. We thus agree with the conclusions that are commonly drawn in the literature about the origin of fossil groups \cite[e.g.,][]{dariush_etal07}, and we have extended these results to the full magnitude gap distribution, rather than just its high-gap tail.

We have generalized these data randomization techniques to
test an underlying assumption that is made in common implementations of the Conditional Luminosity Function (CLF) formalism.
In constructing mock catalogs of galaxies from an N-body simulation with the CLF, each host halo of mass $\mh$ is associated with a ``central galaxy" whose brightness is drawn at random from a log-normal luminosity function, $\phicen(L|\mh),$ with mean denoted by $L_{c}(\mh).$ The brightness of satellite galaxies is presumed to be drawn from an independent distribution, $\phisat(L|\mh),$ which is conventionally modeled as a modified Schechter function that is, in principle, conditioned only by $\mh.$ In practice, however, $\phisat(L)$ is commonly assumed \citep[e.g.,][]{cacciato_etal12,more_etal12,vandenbosch_etal12} to be sufficiently conditioned by the brightness of the central galaxy, so that $L_{c}(\mh)$ effectively determines the host halo mass-dependence of the parameters specifying the brightness distribution of satellite galaxies.

The second data randomization we pursued provides a direct test of the sufficiency of the assumption that the conditioning of $\phisat(L)$ can be adequately encoded by $L_{c}(\mh).$ We accomplish this by constructing mock realizations of group galaxies by using the luminosity of the brightest group member as $L_{c},$ and drawing brightnesses for the remaining members, the satellites, at random from the set of all galaxies that are found in groups whose brightest member is within $0.2$dex of $L_c.$ This mimics the CLF construction method described above, but is more general because we do not assume an analytical form for the luminosity functions of either central or satellite galaxies. Thus this randomization is predicated upon two assumptions, 1) $\Phi_{cen}(L)\approx\Phi(L_1),$ that is, the luminosity distribution of central galaxies, in the sense of the CLF formalism, is accurately approximated by that of the brightest group galaxies, and 2) that $L_{cen}$ is the only property of a galaxy group that is required to describe the brightness distribution of the group's satellites. 

One possible explanation for the good agreement between the predicted and randomized $\monetwo$ distributions shown with the red points in Figure \ref{fig:psigap} is that one or both of the assumptions underlying the data randomization is incorrect, but that $\Phi(\monetwo)$ is a poor statistic to use to discriminate between competing models of the imprint of group assembly on  galaxy brightness. However, the results of our first data randomization, plotted in blue diamonds, argue against this interpretation. With our measurement of $\Psi(\monetwo|>N)$ in the first data randomization we have unambiguously ruled out the hypothesis that group galaxy brightnesses are drawn from a universal luminosity function, demonstrating that the gap distribution can, indeed, provide valuable information about how galaxies of a given brightness are arranged into groups. 

We argue in favor of a second interpretation: {\em knowledge of the luminosity of the central galaxy is sufficient to accurately describe the brightness distribution of satellite galaxies}. As far as we know, our confirmation of this underlying assumption of the CLF formalism is the first of its kind and is unique in at least two respects. First, our test makes no assumptions whatsoever about the functional form of $\Phi_{sat}(L),$ but relies on the data itself to determine the precise form.  Second, we make no use of an N-body simulation, making our test completely independent of any possible systematic errors related to, for example, halo-finding or the fiducial cosmological parameter set of the simulation. Our results therefore provide very general support for this common underlying assumption of the CLF. In closing, we note that we have demonstrated the success of this formalism by considering $\monetwo,$ although other analogous tests are possible. For example, we have not presented results concerning $\Psi(\li),$ $i\geq2,$ nor alternative, gap-based statistics $\Psi(m_{\mathrm{ij}});$ we intend to further develop our data randomization methodology, exploring these and other statistics, in future work.

%--------------------------
\section{Brief Summary of Results}
\label{section:conclusion}
%---------------------------

We have used the SDSS Mr19 catalog of galaxy groups to provide a
series of new tests of subhalo abundance matching (SHAM) models for the connection between galaxies and
dark matter halos. We conclude this paper with a brief summary of our
primary results.

\ben
\item We have developed a novel implementation of SHAM that allows for
the rapid construction of a mock galaxy catalog with a brightness
distribution that {\em exactly matches} any desired luminosity
function, {\em even after scatter has been included.} 
\item Our fiducial SHAM model, based on abundance matching on $\vacc$
with $0.2$dex of scatter at the faint end and $0.15$dex at the bright
end, accurately predicts the group multiplicity function, the abundance of groups
as a function of richness, $g(N),$ a new success for the abundance
matching prescription.
\item The $g(N)$ predictions based on SHAM models using $\vpeak$
and $\vzero$ do not match the observed group multiplicity function.  
In fact, these predictions straddle the $\vacc$ prediction, 
so measurements of group multiplicity may provide 
a promising avenue for constraining models of
satellite mass stripping.
\item No SHAM model studied in either this work or \citet{reddick_etal12}
 can simultaneously account for the observed group multiplicity function and  two-point projected galaxy clustering measurements.
\item The group galaxy luminosity function $\Phi_{group}(L)$ and field
galaxy luminosity function $\Phi_{field}(L)$ are predicted rather
poorly by our mock catalogs, with SHAM group galaxies being
systematically too bright and SHAM field galaxies systematically too
dim. Since our all-galaxy luminosity function exactly matches that of
the observed catalog by construction, this shortcoming must be due to
an erroneous allocation of galaxies into group and field environments. 
We find this to be true in all of the variations of SHAM
catalogs that we explored, suggesting that {\em this is a generic
weakness of the SHAM prescription.}
\item Our fiducial SHAM model, as well as models using $\vpeak$ and
$\vzero$ with the same amount of scatter, accurately predicts the
observed abundance of groups as a function of magnitude gap,
$\Phi(\monetwo),$ suggesting that the prediction for the relative 
brightnesses of galaxies in groups is a new success of the SHAM
paradigm.
\item The gap abundance prediction is quite sensitive to the amount of
scatter between luminosity and $\vmax,$ suggesting that
$\Phi(\monetwo)$ measurements may be a new way to constrain the
scatter in abundance matching.
\item The observed gap abundance is inconsistent with the hypothesis
that the gap is determined by a set of random draws from a universal
luminosity function, contradicting recent claims in the literature \citep{paranjape_sheth11}.
\item The hypothesis that satellite galaxy brightnesses are drawn at
random from $\Phi_{sat}(L|L_{cen})$ is well-supported by observations of magnitude gaps in the SDSS DR7 
groups. We have demonstrated this in a way that is independent from any assumptions concerning the analytic form of  $\Phi_{sat}(L|L_{cen}),$ and without any appeal to numerical simulations. We thus provide very general observational support for this common assumption of the CLF formalism.
\een

\section*{acknowledgments}

We have benefitted greatly from discussions with 
Nick Battaglia, Surhud More, Maya Newman, and Andrey Kravtsov. 
We thank Risa Wechsler and Rachel Reddick for helpful discussions concerning the comparisons of our results to theirs, and Ravi Sheth for comments on an early draft of this paper. We also thank Peter Behroozi for helping us to correctly use and understand the Rockstar halo catalogs.  

The MultiDark Database used in this paper and the web application providing online access to it were constructed as part of the activities of the German Astrophysical Virtual Observatory as result of a collaboration between the Leibniz-Institute for Astrophysics Potsdam (AIP) and the Spanish MultiDark Consolider Project CSD2009-00064. The Bolshoi and MultiDark simulations were run on the NASA's Pleiades supercomputer at the NASA Ames Research Center.

APH and ARZ were supported in part by the Pittsburgh 
Particle physics, Astrophysics, and Cosmology Center (PITT PACC)
at the University of Pittsburgh. 
APH is also supported by the U.S. Department of Energy under contract No. DE-AC02-07CH11359.
The work of ARZ is also 
supported in part by the National Science Foundation 
through grant AST 1108802.  AAB is supported by the Alfred P. Sloan Foundation and by the National Science Foundation through grant AST-1109789.
JAN is supported by the United States Department of Energy Early Career program via grant DE-SC0003960 and NSF AST grant 08-06732.

\bibliography{pgap}

%--------------------------
\section*{Appendix A: SHAM Method}
\label{section:appendix}
%---------------------------

In this Appendix we give a detailed account of our implementation of
the abundance matching procedure to assign galaxies with r-band
luminosities to dark matter halos. As discussed in
\S~\ref{section:mocks}, a mapping from the maximum circular velocity of a
halo, $\vmax,$ to an r-band luminosity $\mr$ is provided by the
implicit relation given by Eq.~\ref{eq:lv}. As we demonstrate in \S~\ref{section:prediction}, different choices for the abundance matching parameter (that is, $\vzero,$ $\vacc,$ and $\vpeak$) result in mock galaxy catalogs with different properties, and so this choice has important consequences in the modeling of the galaxy-halo connection. We remind the reader that we denote mock catalogs constructed by abundance matching with $\vzero$ as ``SHAM0", with $\vacc$ as ``SHAMacc", and with $\vpeak$ as ``SHAMpeak", and that the mock catalog referred to in the text as our fiducial model is a SHAMacc catalog. Since the novel features of our SHAM implementation method are the same regardless of this choice, throughout this Appendix we simply refer to the abundance matching parameter as $\vl.$

Our SHAM procedure begins by using Eq.~(\ref{eq:lv}) to match the
distribution of luminosities assigned to dark matter halos and
subhalos to the double-Schechter function fit in
\citet{blanton_etal05}. We refer to these luminosities as $\linit.$
SHAM models with scatter between $\vl$ and $\mr$ more successfully
describe a variety of astronomical data \citep[see][and references
therein]{klypin_etal11,trujillo-gomez_etal11,watson_etal11} than
models with no scatter. Accordingly, we introduce scatter as follows. For the
$\ith$ halo in the catalog, we assign an independently chosen random
variable $\delta\mri$ drawn from a Gaussian distribution of width
$\sigmai.$ We use these random variables to assign new luminosities to
the galaxies in the catalog via $\mr^{i,\mathrm{init}}\rightarrow\mr^{i,\mathrm{init}}+\delta\mri.$ In
our fiducial catalog, we choose $\sigmai=0.5$ for all halos, which
introduces roughly $0.2$dex of scatter in the galaxy luminosities at the faint end of the luminosity function and $0.15$dex at the bright end,
which is very similar to the level of scatter used in \citet{trujillo-gomez_etal11}. We refer to these brightnesses as
$\lscatter.$

Our goal is to construct a mock catalog with a luminosity function
that exactly matches that of the Mr19 catalog, rather than the
\citet{blanton_etal05} luminosity function. To accomplish this, we
rank-order all the halos and subhalos in the simulation by their
luminosities $\lscatter.$ Because of the scatter we have introduced, this ordering of the halos is
non-monotonic in $\vl.$ 

Rank-ordering the observed Mr19 galaxies by their luminosity naturally provides a map from cumulative number density $\ngal(<\mr)$ to $\mr.$ We use this mapping to associate r-band magnitudes to halos
in Bolshoi. The $\ith$ halo in the list, ordered as described above,
is assigned a rank-ordered cumulative number density $\nrank\equiv i/\vbolshoi,$ where
$\vbolshoi=(250\hmpc)^3.$ We use $\nrank$ to
assign luminosities to the halos by linear interpolation of the map
from $\ngal(<\mr)$ to $\mr.$ Halos with rank-ordered cumulative number
densities larger than $\ngal(<\mr=-19)$
are discarded.\footnote{There are only two out of $\sim10^5$ Bolshoi halos with rank-ordered cumulative 
number densities less than the value of $\ngal$ of the brightest Mr19
galaxy. These halos are not reassigned a new luminosity, but keep
the $\lscatter$ value assigned to them by the initial (post-scatter)
abundance match to the \citet{blanton_etal05} luminosity function.} This procedure gives a
luminosity function of the mock galaxies that {\em exactly matches}
the Mr19 luminosity function, and which includes scatter in the
mapping between $\vl$ and $\mr.$ The reason for the initial abundance match to the \citet{blanton_etal05} analytical fit is simply that the exact luminosity function of galaxies dimmer than $\mr=-19$ that are located in the spatial region occupied by the Mr19 galaxies is not known.

%--------------------------
\section*{Appendix B: Effect of Fiber Collisions on Magnitude Gap Measurements}
\label{section:appendixb}
%---------------------------

Fiber collisions often occur when two or 
more galaxies are located within an angular separation of $55$
arcseconds from one another.  This is the minimum angular separation permitted by the plugging mechanism of the optical fibers used 
in the SDSS spectral measurements. When this occurs,
the fiber is positioned to measure the spectrum of a randomly chosen
galaxy from the two or more ``fiber-collided'' galaxies. 

In this appendix, we discuss how the treatment of fiber collisions influences magnitude gap statistics. There are two separate issues that are relevant to this discussion. First, one must decide whether or not fiber-collided galaxies should be included in the $\monetwo$ measurement of a galaxy group. Second, one must decide what brightness is assigned to the fiber-collided galaxies. We argue below that, for the purpose of measuring magnitude gaps, one should {\em either} 
(1) exclude fiber-collided galaxies from the $\monetwo$ definition, {\em or} (2) include fiber-collided galaxies, but assign them r-band absolute magnitudes according to the prescription we adopt in the catalog used in this paper. Thus when only a catalog with a treatment of fiber collisions that differs from ours is available (as was the case for \citet{paranjape_sheth11}, hereafter PS12), one must exclude fiber-collided galaxies in order to obtain unbiased $\monetwo$ measurements. Below we present our supporting argument for these recommendations. Throughout this appendix we focus on comparing the fiber collision methodology used in our paper to that in PS12 because we reach different conclusions than they do based on the same statistical test, and we trace this difference to fiber collisions. 

In the DR3-based group catalog used in PS12, the remaining galaxies of a fiber-collided set are 
assigned the redshift {\em and} brightness of the randomly-chosen galaxy.  
In the DR7-based group catalog that we utilize in this study, 
only the redshift of the randomly-chosen galaxy is assigned to the
remaining fiber-collided galaxies.  The absolute r-band magnitudes of 
the remaining galaxies in a fiber-collided set are inferred from their apparent 
r-band magnitudes using the redshift of the randomly-chosen, 
spectroscopically-observed galaxy. For convenience, we will refer to the choice to assign {\em both} the redshift {\em and} the r-band absolute magnitude of the spectroscopically measured galaxy to the fiber-collided galaxy as ``the DR3 treatment of fiber collisions", and the choice to assign only the redshift of the observed galaxy to the fiber-collided galaxy as ``the DR7 treatment". We emphasize, however, that this is a post-processing choice and that either data set could be treated in either fashion.
 
The magnitude gap measurements presented in PS12 differ from ours in two important respects. First, we define $\monetwo$ to be the r-band magnitude difference between the two brightest {\em non-fiber-collided} members of the group, whereas PS12 use fiber-collided galaxies in their $\monetwo$ definition. Second, the catalog used in PS12 employs the DR3 treatment of fiber collisions, while we use a catalog based on the DR7 treatment. 

When using a catalog with the DR3 treatment of fiber collisions {\em and} including fiber-collided galaxies in the definition of $\monetwo,$ all groups with a member that is fiber-collided with the brightest member are assigned magnitude gaps precisely equal to zero. This is a relatively common scenario since brightest group galaxies are typically found in the densest regions of the sky. Thus with these conventions there is an artificial ``spike" in the magnitude gap distribution at $\monetwo=0.$ In our magnitude gap measurements, the $\monetwo=0$ spike does not occur.\footnote{Although we note that
when we run our analysis pipeline on the DR3 data set and include fiber-collided galaxies in the $\monetwo$ measurement, we recover the \citet{paranjape_sheth11} results in full quantitative detail.} In addition to incorrectly enhancing the abundance of $\monetwo=0$ groups, this treatment of fiber collisions also results in an abundance of large- and moderate-gap groups that is systematically too low, since $100\%$ of such groups with a galaxy that is fiber-collided with the brightest member are incorrectly removed from large- and moderate-gap bins and assigned $\monetwo=0.$ 
%As one can see by the sign of the blue diamonds in $\Psi(\monetwo|>N)$ in Fig.~\ref{fig:psigap}, the sense of this systematic error is such that it improves the agreement between $\Phi(\monetwo|> N)$ and its Monte Carlo relative to our treatment.

%---------------------------------------------------------------------------------------------------
\begin{figure}
\centering
\includegraphics[width=8.0cm]{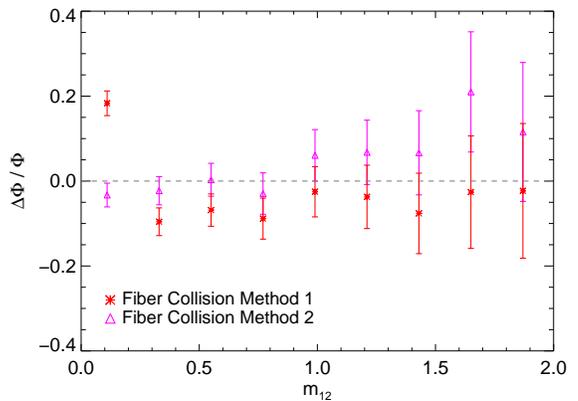}
\caption{
Monte Carlo test of the influence of fiber collisions on the magnitude gap distribution, $\Phi(\monetwo).$ The fractional difference between the ``true" (Monte Carlo) and ``measured" gap abundance is plotted for two different methods of treating fiber collisions. Method 1, plotted with red asterisks, includes fiber-collided galaxies in the $\monetwo$ measurement, and employs ``the DR3 treatment" of fiber collisions, resulting in an erroneous ``spike" at $\monetwo=0$ that biases the gap measurement. The Method 1 measurement of $\Phi(\monetwo)$ differs from the true distribution at $4.8\sigma.$ Method 2, our method, plotted in magenta triangles, excludes fiber-collided galaxies in measuring $\monetwo$ and faithfully recovers the true gap distribution. 
}
\label{fig:fctest}
\end{figure}
%-----------------------------------------------------------------------------------------------------

We have performed a simple test of our claim that excluding fiber-collided galaxies from the $\monetwo$ definition yields unbiased magnitude gap measurements. We proceed as follows. Denoting the luminosity function exhibited by the galaxies in the Mr19 galaxy group catalog by $\phiglo(L),$ we begin by assigning new r-band magnitudes to each galaxy in the catalog by randomly drawing luminosities from $\phiglo(L).$ These randomly selected luminosities will be treated as the ``true" brightnesses throughout this exercise. In this fashion, we construct a new, Monte Carlo group catalog in which the galaxies have randomly selected brightnesses, but we retain the information from the Mr19 catalog about each galaxy's group membership and fiber collision (where relevant). To be clear, nowhere in this exercise do we use the measured brightnesses of the galaxies (except , of course, in the construction of $\phiglo(L).)$

For each group in the Monte Carlo catalog, we measure the magnitude gap $\monetwo$ in three different ways:
\ben
\item We use each galaxy's randomly assigned magnitude to measure $\monetwo,$ {\em including} the fiber-collided members of the group. The resulting gap distribution $\Phi(\monetwo)$ is treated as the ``true" distribution since in this exercise the randomly assigned brightnesses are the true brightnesses. 
\item As above, we include all members of the group in the $\monetwo$ measurement, but only after first assigning to each fiber-collided member the luminosity of its non-fiber-collided counterpart. This mirrors the DR3 treatment of fiber collisions together with the definition for $\monetwo$ adopted by \citet{paranjape_sheth11}. We will refer to this as ``Fiber Collision Method 1".
\item We use each galaxy's randomly assigned magnitude to measure $\monetwo,$ {\em excluding} the fiber-collided members of the group. This mirrors the methodology we adopt in this paper. We will refer to this as ``Fiber Collision Method 2".
\een

In Figure \ref{fig:fctest}, we plot the fractional difference between the true (Monte Carlo) $\Phi(\monetwo)$ and that which is measured by the alternate methods (ii) and (iii), described above. Results pertaining to the $\monetwo$ measurement using Fiber Collision Method 1 appear in red asterisks; results obtained via Fiber Collision Method 2 are plotted in magenta triangles. 

In the lowest $\Phi(\monetwo)$ bin, the effect of the $\monetwo=0$ ``spike" that occurs in Method 1 is quite visible. The $\Phi(\monetwo)$ measurement obtained via Method 1 differs from the true (Monte Carlo) distribution at a level of $4.8\sigma.$ On the other hand, there is virtually zero statistically significant difference between the Method 2 measurement of $\Phi(\monetwo)$ and the true (Monte Carlo)  distribution. Moreover, it is also evident that the $\monetwo=0$ spike causes $\Phi(\monetwo)$ to be systematically too low in {\em all} of the other bins. Even when the lowest $\monetwo$ bin is excluded from the $\chi^2$ test, the difference between the true and Method 1-measured $\Phi(\monetwo)$ persists at a level of $2\sigma.$ 

%{\em Using our DR7 treatment of fiber collisions, the gap abundance is the same regardless of whether or not fiber-collided galaxies are used in the definition of $\monetwo,$ whereas with the DR3 treatment used in \cite{paranjape_sheth11} the gap abundance measurement is very sensitive to this choice.}  Moreover, using our $\monetwo$ definition the gap abundance exhibited in the DR7 data set (using the DR7 fiber collision treatment) is virtually identical to the gap abundance exhibited in the DR3 data set (using the DR3 fiber collision treatment). On the other hand, using the $\monetwo$ definition adopted in \citet{paranjape_sheth11} these two gap abundances are very different from one another.

Our Monte Carlo exercise demonstrates the following: measuring the
magnitude gap of a galaxy group from the non-fiber-collided members
yields an unbiased measurement of the group's true $\monetwo.$ This
justifies the methodology we have adopted in this paper. The MC
additionally shows that, when including fiber-collided galaxies in
$\monetwo$ measurements, the DR3 treatment of fiber collisions results
in biased measurements of the magnitude gap. These differences
entirely account for the differences between the conclusions of this
paper and those drawn in PS12. In particular, with our unbiased magnitude gap measurements we conclude that the global distribution $\Phi(\monetwo)$ is inconsistent with the distribution that results from random drawing galaxy brightnesses from a universal luminosity function (see \S~\ref{section:mcs}). 

We note that \citet{paranjape_sheth11} were careful to point out that their tests based on two-point statistics were inconsistent with their conclusions concerning the random draw hypothesis. Our results in this appendix demonstrate that fiber collisions were the culprit for this inconsistency. 

We conclude this appendix by pointing out that it {\em is} possible to include fiber-collided galaxies in $\monetwo$ measurements and still obtain unbiased results, but one must use a catalog that is based on the DR7 treatment of fiber collisions, not the DR3 treatment. The evidence for this is simple: using our Mr19 SDSS group catalog, which employs the DR7 treatment, we measure same distribution $\Phi(\monetwo)$ whether or not we include fiber-collided galaxies in our gap measurements. Since we have already shown that excluding fiber-collided members results in unbiased $\monetwo$ measurements, it follows that if fiber-collided galaxies are assigned luminosities according to the DR7 treatment, including these galaxies in gap measurements also results in an unbiased $\monetwo.$ A group catalog constructed from SDSS DR3 data that is based on the DR7 treatment of fiber collisions can be found at   
http://lss.phy.vanderbilt.edu/groups. An update to this catalog that is based on DR7 data will be presented in a forthcoming paper.

%http://cosmo.nyu.edu/aberlind/Groups
%--------------------------
\section*{Appendix C: Tests of Multiplicity Function Systematics}
\label{section:appendixc}
%---------------------------

One of the primary conclusions of \S~\ref{subsection:gn} is that SHAMpeak models overestimate the abundance of groups with $N\gtrsim20$ members. In this appendix we explore the influence of two possible systematics that may influence this result. First, effects due to the edges of the SDSS survey that are not present in the periodic Bolshoi box may affect measurements of group multiplicity. If the centroid of a group of galaxies happens to be near an edge of the survey volume of our Mr19 SDSS galaxy sample, the galaxies beyond the survey edge will not be included as group members. This would artificially deplete the richness of all such groups.

To estimate the magnitude of the influence of edge effects, we have excluded groups whose center is within $1\hmpc$ of the nearest survey edge and recomputed $g(N).$ This cut excludes $14\%$ of the groups in our sample, and so we estimate the effective volume of this reduced sample as being $86\%$ of the original survey volume. We find that the differential multiplicity function of the reduced sample is within $1\sigma$ of the $\dd g(N)/\dd\log N$ of the original sample, with fractional differences of $\Delta\dd g/\dd\log N<18\%$ in all richness bins. Meanwhile, the SHAMpeak discrepancy for the multiplicity function prediction (see Fig.~\ref{fig:gn}) differs at the level of $50-80\%$ from the observed $g(N).$ We conclude that it is unlikely that edge effects account for the erroneous prediction of the multiplicity function in SHAMpeak models. 

A second possible systematic in our analysis could come from fiber collisions. In our mock galaxy sample, we have only introduced fiber collisions after identifying mock groups. Our procedure was designed to account for the influence of fiber collisions on magnitude gaps, but it does not account for any possible influence that fiber collisions have on the multiplicity function. 

We estimate the impact that fiber collisions could have on the group multiplicity function by first excluding fiber-collided galaxies from our Mr19 SDSS galaxy sample and then identifying groups on the reduced set of galaxies. As $\sim5\%$ of galaxies in the Mr19 sample are fiber-collided, we estimate the effective volume of the reduced data set as $95\%$ of the original effective volume.  In this case, the fractional differences are at a level of $\Delta\dd g/\dd\log N<15\%$ in all richness bins. We similarly conclude that it is unlikely that fiber collisions account for the erroneous prediction of the multiplicity function in SHAMpeak models.

\end{document}